\documentclass[sn-mathphys-num]{sn-jnl}% Math and Physical Sciences Numbered Reference Style 
\usepackage{graphicx}%
\usepackage{multirow}%
\usepackage{amsmath,amssymb,amsfonts}%
\usepackage{amsthm}%
\usepackage{mathrsfs}%
\usepackage[title]{appendix}%
\usepackage{xcolor}%
\usepackage{textcomp}%
\usepackage{manyfoot}%
\usepackage{booktabs}%
\usepackage{algorithm}%
\usepackage{algorithmicx}%
\usepackage{algpseudocode}%
\usepackage{listings}%
\theoremstyle{thmstyleone}%
\theoremstyle{thmstyletwo}%

\theoremstyle{thmstylethree}%

\begin{document}

\title[Multivariate Bernoulli Subcopula]{A subcopula characterization of dependence for the Multivariate Bernoulli Distribution}


\author*[1]{\fnm{Arturo} \sur{Erdely}}\email{aerdely@acatlan.unam.mx}



\affil*[1]{\orgdiv{Facultad de Estudios Superiores Acatl\'an}, \orgname{Universidad Nacional Aut\'onoma de M\'exico}, \orgaddress{\street{Av. Alcanfores y San Juan Totoltepec S/N}, \city{Naucalpan}, \postcode{53150}, \state{Estado de M\'exico}, \country{M\'exico}}}

%\affil[2]{\orgdiv{Department}, \orgname{Organization}, \orgaddress{\street{Street}, \city{City}, \postcode{10587}, \state{State}, \country{Country}}}

%\affil[3]{\orgdiv{Department}, \orgname{Organization}, \orgaddress{\street{Street}, \city{City}, \postcode{610101}, \state{State}, \country{Country}}}


\abstract{%% By applying Sklar’s theorem to the Multivariate Bernoulli Distribution, this paper proposes a framework to decouple marginal distributions from the dependence structure, clarifying interactions among binary variables. 
This paper provides a subcopula characterization of dependence for the Multivariate Bernoulli Distribution. Explicit formulas are derived using subcopulas to introduce dependence measures for interactions of all orders, not just pairwise. A Bayesian inference approach is also applied to estimate the parameters, offering practical tools for parameter estimation and dependence analysis in real-world applications. %%The results obtained contribute to the application of subcopulas of multivariate binary data, with real data examples. 
The main contribution is an explicit, order-by-order characterization of multivariate dependence for binary data (beyond pairwise association), linking joint probabilities to subcopula-based dependence parameters/measures under full compatibility; this fills a gap in the literature where practical modeling often stops at pairwise dependence.
}

\keywords{Sklar's theorem, subcopula, dependence, Bernoulli, binary data}



\maketitle

\section{Introduction}\label{sec:intro}

In statistical modeling, the analysis of multivariate binary data often requires flexible methods to capture the dependence structure between variables. The Multivariate Bernoulli Distribution (MBD) offers a natural framework for modeling such data, yet specifying and understanding the dependence between variables in this context can be challenging. This is especially true when working with multivariate data where traditional correlation measures fall short due to the binary nature of the variables and the dimensionality.

A representation of the MBD that provides an alternative to the traditional log-linear model for binary variables was proposed by \cite{bib1} using the concept of Kronecker product from matrix calculus, in terms of $2^n-1$ parameters, where $n$ is the dimension of a random vector with MBD, but where the dependency vector of parameters is, in fact, non standardized central moments. In \cite{bib2} several flexible methods for simulating random binary sequences with fixed marginal distributions and specified degrees of association between the variables are discussed, but only pairwise dependencies are considered. An algorithm is derived in \cite{bib3} for generating systems of correlated binary data, but allowing for the specification of just pairwise correlations within each system. A simple method to characterize multivariate Bernoulli variables with given means is investigated in \cite{bib4} but only pairwise correlations are discussed.

Copulas, and more generally subcopulas, provide a straightforward approach to describing the dependence structure in multivariate settings by decoupling the marginal distributions from the joint distribution. Introduced by Sklar's theorem \cite{bib5}, subcopulas extend copula theory to cases where the marginals are not necessarily continuous, making them particularly useful for multivariate discrete data. 

%% Response to Editor comment 5)
From a computational perspective, subcopulas arise naturally on the discrete grid of attainable marginal cdf values, and they are the relevant objects for algorithmic work with discrete data, see \cite{Nguyen2019}. In this sense, the proposed explicit subcopula-based formulas and compatibility-preserving parameter selection procedures provide directly implementable building blocks for dependence modeling and inference with multivariate binary data.

%%%%% In response to Editor 2)
Copula models have become standard tools across a broad range of applied fields (energy, environmental sciences, forestry, hydrology, finance, insurance, among others), largely because they separate marginal modeling from dependence modeling in a flexible way; see \cite{Bhatti2019} for a wide-ranging review and applications. 
%%%%% In response to Editor 6)
In finance, copula and EVT-based dependence modeling has also been used to study tail dependence and directional/causal dependence in markets, see for example \cite{AlRahahleh2017}.
%%%%% In response to Editor 7)
Recent work also uses copula-based contagion frameworks to study global shock transmission and tail-risk spillovers, see \cite{Saekow2025}.

%%%%%%% In response to reviewer 1 %%%%%%%%
In \cite{bib5A} it is discussed that many of the convenient properties of copulas do not carry over from the continuous to the discrete case, because the underlying copula is not unique, but according to \cite{bib5} there is a unique subcopula, and that is the reason why this paper focuses on leveraging subcopulas to characterize the dependence within the MBD, offering a framework to better understand and quantify dependence of all orders (pairwise, three-wise, four-wise, etc.) 

%%%%% Response to Editor 8)
In particular, \cite{vanEs2025} emphasize that with discrete margins the copula is generally non-unique and likelihood-based inference relies on inclusion–exclusion rectangle probabilities rather than densities (and may be affected by ties), proposing tau-informed initialization and computationally efficient ML estimation for Gaussian copulas with discrete data. The present contribution is complementary by directly working with the unique subcopula induced by Multivariate Bernoulli data and provides an explicit, compatibility-preserving characterization of dependence parameters/measures across all interaction orders for binary margins. The present work introduces a formal subcopula-based characterization of the dependence structure within the MBD, and illustrates its usefulness to tackle the compatibility problem of building multivariate models in terms of lower dimension marginals.

This work provides a systematic approach for analyzing the joint dependence of multivariate binary data, with potential applications in fields ranging from genetics to econometrics where binary outcomes are prevalent. The results presented here not only contribute to the theoretical understanding of subcopulas but also offer practical tools for applied statistics and data analysis, and Julia programming \cite{bib6} code is provided for their computational implementation and application.

\section{Subcopulas and dependence}\label{sec:subcopula}

Sklar's Theorem \cite{bib5} proves that for any $n-$dimensional vector of random variables $(X_1,\ldots,X_n)$ there exists a functional link $S^{(n)}$ between its joint probability distribution $F_{X_1,\ldots,X_n}(x_1,\ldots,x_n)=\mathbb{P}(X_1\leq x_1,\ldots,X_n\leq x_n)$ and its corresponding univariate marginal distributions $F_{X_i}(x_i)=\mathbb{P}(X_i\leq x_i),$ $i\in\{1,\ldots,n\}:$
\begin{equation}\label{sklar}
	F_{X_1,\ldots,X_n}(x_1,\ldots,x_n) \,=\, S^{(n)}\left(F_{X_1}(x_1),\ldots,F_{X_n}(x_n)\right)\,,
\end{equation}
for any  $(x_1,\ldots,x_n)\in\overline{\mathbb{R}}^n,$ and where the function $S^{(n)}: D_1\times\cdots\times D_n\rightarrow[0,1]$ is unique, with $D_i=\mbox{Ran}\,F_{X_i}=\{F_{X_i}(x):x\in\overline{\mathbb{R}}\}$ for $i\in\{1,\ldots,n\},$ which implies $\{0,1\}\subseteq D_i\subseteq [0,1].$ Such link $S^{(n)}$ is called a \textit{subcopula} function, and in the particular case where every $D_i=[0,1]$ then it is called a \textit{copula} function, which would be the case when all the random variables are continuous. If at least one of the random variables is not continuous then the domain of the underlying subcopula $S^{(n)}$ is a proper subset of the unit hypercube $[0,1]^n.$ A formal definition for subcopula functions and their properties are discussed in detail in \cite{bib7} and \cite{bib8}.

Recalling that the random variables $X_1,\ldots,X_n$ are mutually independent if and only if their probability joint distribution is equal to the product of its univariate marginals, then as an immediate consequence of (\ref{sklar}) the unique subcopula function under independence is:
\begin{equation}\label{indep}
	\Pi^{(n)}(u_1,\ldots,u_n) := u_1\cdots u_n\,,\qquad (u_1,\ldots,u_n)\in D_1\times\cdots\times D_n\,.
\end{equation}

Moreover, as a consequence of the Fr\'echet-Hoeffding bounds for joint distributions (see \cite{bib9} and \cite{bib10}) and, applying (\ref{sklar}), we may obtain bounds for any subcopula $S^{(n)}:$
\begin{equation}\label{FH}
	W^{(n)}(u_1,\ldots,u_n) \,\leq\, S^{(n)}(u_1,\ldots,u_n) \,\leq\, M^{(n)}(u_1,\ldots,u_n)\,,
\end{equation}
where $W^{(n)}(u_1,\ldots,u_n):=\max\{u_1+\cdots+u_n-n+1,0\}$ and $M^{(n)}(u_1,\ldots,u_n):=\min\{u_1,\ldots,u_n\}.$ The upper bound $M^{(n)}$ is always a subcopula, but the lower bound $W^{(n)}$ is subcopula only for $n=2,$ though it is still the best possible lower bound for $n>2,$ see \cite{bib8}. 

From any $n-$dimensional subcopula $S^{(n)}(u_1,\ldots,u_n)$ and $2\leq k<n$ it is possible to obtain $\binom{n}{k}$ $k-$dimensional \textit{marginal subcopulas} $S^{(k)}(v_1,\ldots,v_k)$ defined in terms of $S^{(n)}(u_1,\ldots,u_n)$ where $k$ of its entries are equal to exactly one of each $\{v_1,\ldots,v_k\}$ and the remaining $n-k$ are equal to $1.$

Since the univariate marginal distributions $F_{X_i}(x_i)=\mathbb{P}(X_i\leq x_i)$ have no information about how each random variable $X_i$ interacts with others, then as a consequence of (\ref{sklar}) all the information about the dependence among the random variables $X_1,\ldots,X_n$ is contained in their unique underlying subcopula $S^{(n)},$ and therefore any attempt to measure degrees or intensity of dependence should extract information from $S^{(n)}.$ 

For the particular case of bivariate copulas (when both random variables are continuous) several copula-based measures have been studied, such as the \textit{concordance} measures by Spearman \cite{bib11} or Kendall \cite{bib12}, or \textit{dependence} measures by Schweizer-Wolff \cite{bib13} or Hoeffding \cite{bib10}. 

%%%% Response to Reviewer 1 %%%%%%
But when at least one of the variables is discrete, as pointed out by \cite{bib5A}, an identifiability issue arises, and therefore copula-based dependence or concordance measures cannot be adapted in this case since it is not possible to report a unique value. This motivated the work by \cite{bib14} where for the case of proper subcopulas (that are not copulas) a subcopula dependence measure has been proposed. For the case of a vector $(X,Y)$ of arbitrary type random variables with underlying and unique subcopula $S$ it is defined:
\begin{equation}\label{eq:mdm2}
	d(S) := \sup_{\text{Dom}\,S}\{S-\Pi_S\} - \sup_{\text{Dom}\,S}\{\Pi_S-S\}\,.
\end{equation}
where for any $(u,v)\in\text{Dom}\,S=\text{Ran}\,F_X\times\text{Ran}\,F_Y$ the following subcopulas are defined as $W_S(u,v):=\max\{u+v-1,0\},$ $M_S(u,v):=\min\{u,v\},$ and $\Pi_S(u,v)=uv,$ and then:
\begin{equation}\label{eq:mdm}
	\mu_{X,Y} \equiv \mu_S := \begin{cases}
		d(S)/d(M_S) & \text{ if } d(S)\geq 0 \text{ and } M_S\neq\Pi_S, \\ 
		-d(S)/d(W_S) & \text{ if } d(S)\leq 0 \text{ and } W_S\neq\Pi_S, \\
		\qquad 0 & \text{ if } W_S = \Pi_S = M_S.
	\end{cases}
\end{equation}
where $d(M_S)$ and $d(W_S)$ is formula (\ref{eq:mdm2}) applied to $M_S$ and $W_S,$ respectively. In (\ref{eq:mdm2}) it is calculated the largest difference between the unique underlying subcopula $S$ and the subcopula $\Pi_S$ that represents independence, and this value is used in (\ref{eq:mdm}) to calculate a proportion with respect to the largest possible differences, according to the Fr\'echt-Hoeffding bounds for subcopulas. In fact, (\ref{eq:mdm2}) and (\ref{eq:mdm}) may be similarly applied to any subcopula in higher dimensions, using (\ref{indep}) and (\ref{FH}).

\section{Bivariate Bernoulli}\label{sec:bivariate}

Let $(X_1,X_2)$ be a vector of Bernoulli random variables with (univariate marginal) parameters $1-\theta_r$ where $0<\theta_r<1,$ that is $\mathbb{P}(X_r=0)=\theta_r$ and $\mathbb{P}(X_r=1)=1-\theta_r$ for $r\in\{1,2\}.$ Then the univariate marginal distribution functions are given by $F_r(x)=\mathbb{P}(X_r\leq x)=\theta_r\mathbb{I}_{\{0\,\leq\,x\,<\, 1\}} + \mathbb{I}_{\{x\,\geq\,1\}},$ where $\mathbb{I}_A$ stands for the indicator (or characteristic) function of the set $A,$ and therefore $\mbox{Ran}\,F_r=\{0,\theta_r, 1\}.$ 

According to (\ref{sklar}) the domain of the underlying subcopula in this case is the set $\mbox{Dom}\,S^{(2)}=\{0,\theta_1,1\}\times\{0,\theta_2,1\}.$ From the formal definition of bivariate subcopula (see \cite{bib8}, for example) it always satisfies $S^{(2)}(u,0)=0=S^{(2)}(0,v),$ $S^{(2)}(u,1)=u,$ and $S^{(2)}(1,v)=v,$ and therefore:
\begin{equation}\label{subcop2d}
	S^{(2)}(u,v) = \begin{cases}
		\,1 & \mbox{ if } (u,v)=(1,1), \\
		\,\theta_1 & \mbox{ if } (u,v)=(\theta_1,1), \\
		\,\theta_2 & \mbox{ if } (u,v)=(1,\theta_2), \\
		\,\theta_{12} & \mbox{ if } (u,v)=(\theta_1,\theta_2), \\
		\,0 & \mbox{ elsewhere,}
	\end{cases}
\end{equation}
where $\theta_{12}$ is any value satisfying:
\begin{equation}\label{FH2d}
	\max\{\theta_1+\theta_2-1,0\}\,\leq\,\theta_{12}\,\leq\,\min\{\theta_1,\theta_2\}
\end{equation}
as a consequence of (\ref{FH}), that in this case acts as a bivariate dependence parameter. Since $\mbox{Ran}\,(X,Y)=\{(0,0), (0,1), (1,0), (1,1)\}$ let $\{p_{00}, p_{01}, p_{10}, p_{11}\}$ be the joint point probabilities:
\begin{equation*}%\label{pij}
	p_{ij} := \mathbb{P}(X_1=i,X_2=j)\,,\qquad i,j\in\{0,1\},
\end{equation*}
where necessarily $0\leq p_{ij}\leq 1$ and $p_{00}+p_{01}+p_{10}+p_{11}=1,$ and therefore only three of the $p_{ij}$ values need to be specified, since the fourth is just $1$ minus the sum of the other three. Applying (\ref{sklar}) and (\ref{subcop2d}) to calculate each $p_{ij}$ in terms of the joint distribution function $F_{12}(i,j)=\mathbb{P}(X_1\leq i,X_2\leq j):$
\begin{eqnarray*}%\label{Bernoulli2dcdf}
	p_{ij} &=& F_{12}(i,j) - F_{12}(i,j-1) - F_{12}(i-1,j) + F_{12}(i-1,j-1)\,, \nonumber \\
	&=& S^{(2)}\left(F_1(i),F_2(j)\right) - S^{(2)}\left(F_1(i),F_2(j-1)\right) - S^{(2)}\left(F_1(i-1),F_2(j)\right) \ldots\nonumber\\
	&{ }& + S^{(2)}\left(F_1(i-1),F_2(j-1)\right)\,,
\end{eqnarray*}
and therefore:
\begin{eqnarray} \label{Bernoulli2dpdf}
	p_{00} &=& F_{12}(0,0) = S^{(2)}(\theta_1,\theta_2) = \theta_{12}\,, \\ 
	p_{01} &=& F_{12}(0,1)-F_{12}(0,0) = S^{(2)}(\theta_1,1)-S^{(2)}(\theta_1,\theta_2) = \theta_1 - \theta_{12}\,, \nonumber \\
	p_{10} &=& F_{12}(1,0)-F_{12}(0,0) = S^{(2)}(1,\theta_2)-S^{(2)}(\theta_1,\theta_2) = \theta_2 - \theta_{12}\,, \nonumber \\
	p_{11} &=& 1 - \theta_1 - \theta_2 + \theta_{12}\,. \nonumber
\end{eqnarray}

Formulas (\ref{Bernoulli2dpdf}) allow us to fully characterize the family of all bivariate Bernoulli distributions in terms of the marginal parameters $0<\theta_1<1$ and $0<\theta_2<1,$ and a bivariate dependence parameter $\theta_{12}$ that must satisfy $\max\{\theta_1+\theta_2-1,0\}\leq\theta_{12}\leq\min\{\theta_1,\theta_2\},$ that we may denote as $\mathcal{B}_2(\theta_1,\theta_2,\theta_{12}).$ The subcopula based dependence measure proposed by \cite{bib14} for $\mathcal{B}_2(\theta_1,\theta_2,\theta_{12})$ becomes:
\begin{equation}\label{depmeas2d}
	\mu(X_1,X_2) =  \begin{cases}
		\frac{\theta_{12}-\theta_1\theta_2}{\min\{\theta_1,\theta_2\}-\theta_1\theta_2} & \mbox{ if } \theta_{12}\geq \theta_1\theta_2 \,, \\
		{ } &     { } \\
		\frac{\theta_{12}-\theta_1\theta_2}{\theta_1\theta_2 - \max\{\theta_1+\theta_2-1,0\}} & \mbox{ if } \theta_{12} <  \theta_1\theta_2 \,.
	\end{cases}
\end{equation}

Two Bernoulli random variables are independent if and only if $\theta_{12}=\theta_1\theta_2$ as an immediate consequence of (\ref{indep}), or equivalently if and only if $\mu(X_1,X_2)=0,$ and therefore:
\begin{equation*}%\label{Bernoulli2dIndep}
	(p_{00},p_{01},p_{10},p_{11})=(\theta_1\theta_2,\theta_1(1-\theta_2),(1-\theta_1)\theta_2,(1-\theta_1)(1-\theta_2))\,,
\end{equation*}
as expected. For two dependent Bernoulli random variables, that is when $\theta_{12}\neq\theta_1\theta_2,$ or equivalently when $\mu(X_1,X_2)\neq 0,$ without loss of generality assume that $\theta_1\leq\theta_2.$ The Fr\'echet-Hoeffding upper bound (\ref{FH}) is attained with dependence parameter $\theta_{12}=\min\{\theta_1,\theta_2\}=\theta_1$ and therefore $\mu(X_1,X_2)=+1$ with:
\begin{equation*}%\label{Bernoulli2dUpper}
	(p_{00},p_{01},p_{10},p_{11})=(\theta_1,0,\theta_2-\theta_1,1-\theta_2)\,.
\end{equation*}
Only the particular case $\theta_1=\theta_2=\theta$ translates into:
\begin{equation*}%\label{Bernoulli2dUpperBis}
	(p_{00},p_{01},p_{10},p_{11})=(\theta,0,0,1-\theta)\,,
\end{equation*}
that is $\mathbb{P}(X_1=X_2)=p_{00}+p_{11}=1.$ Equivalently, if $\theta_1\neq\theta_2$ then it is definitely not possible to get $\mathbb{P}(X_1=X_2)=1.$

The Fr\'echet-Hoeffding lower bound (\ref{FH}) is reached with dependence parameter $\theta_{12}=\max\{\theta_1+\theta_2-1,0\}$ which applied to formulas (\ref{Bernoulli2dpdf}) translates into:
\begin{equation*}%\label{Bernoulli2dLower}
	(p_{00},p_{01},p_{10},p_{11}) = \begin{cases}
		\,(\theta_1+\theta_2-1,1-\theta_2,1-\theta_1,0) & \mbox{if } \theta_1+\theta_2>1\,, \\
		\,(0,\theta_1,\theta_2,1-\theta_1-\theta_2) & \mbox{if } \theta_1+\theta_2\leq 1\,,
	\end{cases}
\end{equation*}
with $\mu(X_1,X_2)=-1,$ from where it is clear that $\mathbb{P}(X_1=1-X_2)=1$ if and only if $\theta_1+\theta_2=1.$

In summary, formulas (\ref{Bernoulli2dpdf}) allow to define the family $\mathcal{B}_2(\theta_1,\theta_2,\theta_{12})$ of all bivariate distributions with Bernoulli univariate marginals, which is in fact also known as a \textit{Fr\'echet class} of the type $\mathcal{F}(F_1,F_2),$ see for example \cite{bib15}. A Fr\'echet class is a set of all multivariate distribution functions that share the same given marginal distributions. Formally, let $F_1,F_2,\ldots,F_d$ be univariate cummulative distribution functions. The Fr\'echet class associated with $(F_1,\ldots,F_d)$ is defined as:
$$\mathcal{F}(F_1,\ldots,F_d):=\{H:H\text{ is a $d$-dimensional distribution with marginals }F_1,\ldots,F_d\}$$

\begin{figure}
	\begin{center} 
		\includegraphics[width=12cm, keepaspectratio]{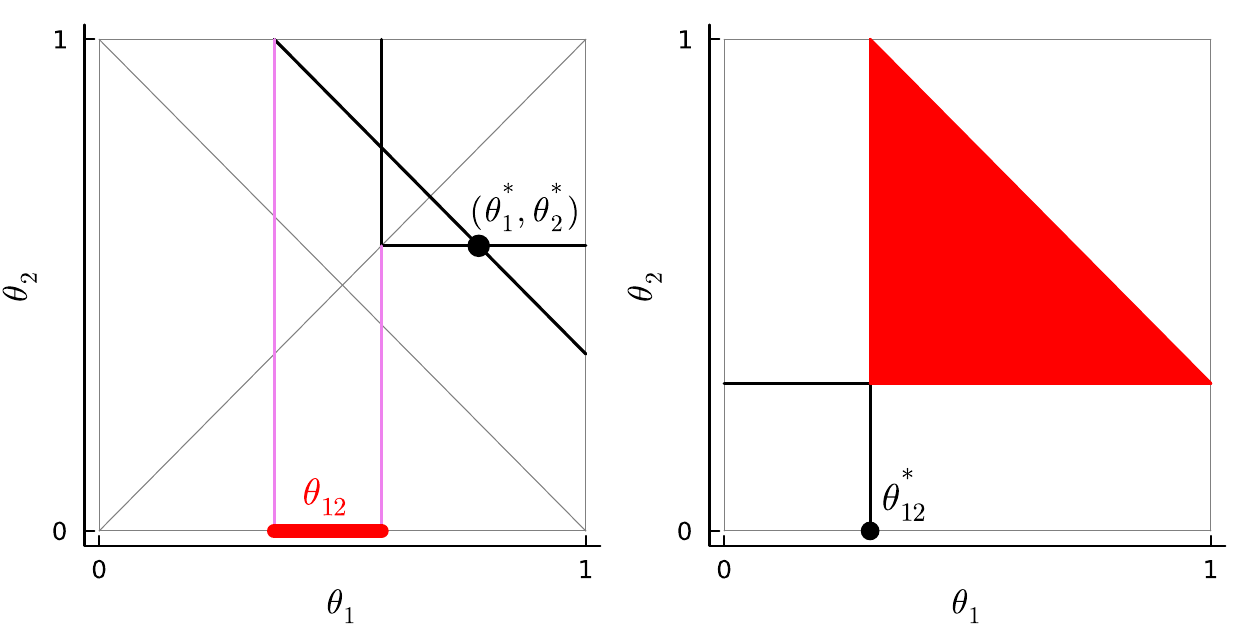}
	\end{center}
	\caption{Bivariate Bernoulli. \textbf{Left}: Compatible values for the bivariate dependence parameter $\theta_{12}$ (red interval), given a specific pair of values $(\theta_1^*,\theta_2^*)$ for the univariate marginal parameters. \textbf{Right}: Compatible values for the marginal univariate parameters $(\theta_1,\theta_2)$ (red triangle region), given a specific value $\theta_{12}^*$ for the bivariate dependence parameter.}
	\label{fig:B2}
\end{figure}

The bivariate joint probability mass function $p_{ij}=\mathbb{P}(X_1=i,X_2=j)$ for $i$ and $j$ in $\{0,1\}$ can be specified in two ways:
\begin{itemize}
	\item[a)] choosing freely the two marginal parameters $0<\theta_1<1$ and $0<\theta_2<1$ and then choosing $\theta_{12}$ such that (\ref{FH2d}) holds, or
	\item[b)] choosing freely the dependence parameter $0<\theta_{12}<1$ and then choosing the marginal parameters $0<\theta_1<1$ and $0<\theta_2<1$ such that (\ref{FH2d}) holds. Equivalently, instead of choosing the dependence parameter value, we may choose a given value of the dependence measure defined by \cite{bib14} and then obtain the value for $\theta_{12}$ from (\ref{depmeas2d}).
\end{itemize}
Behind the above cases a) and b) is what is known as a \textit{compatibility problem,} in this particular case for the Fr\'echet class $\mathcal{F}(F_1,F_2):$ even though the three parameters are the probabilities $\theta_1=\mathbb{P}(X_1=0),$ $\theta_2=\mathbb{P}(X_2=0),$ and $\theta_{12}=\mathbb{P}(X_1=0,X_2=0),$ not all values in the $[0,1]$ interval are simultaneously admissible (or compatible) for such parameters, since (\ref{FH2d}) must hold, see Figure \ref{fig:B2}:
\begin{itemize}
    \item[a)] Figure \ref{fig:B2} Left: Specific values are chosen for $\theta_1=\theta_1^*$ and $\theta_2=\theta_2^*$ and from (\ref{FH2d}) all the admissible values for $\theta_{12}\in\big[\max\{\theta_1^*+\theta_2^*-1,0\}\,,\,\min\{\theta_1^*,\theta_2^*\}\big].$ If the chosen point $(\theta_1^*,\theta_2^*)$ is below the solid gray identity line $\theta_2=\theta_1$ then $\min\{\theta_1^*,\theta_2^*\}=\theta_1^*,$ which is projected to the horizontal axis through such identity line (horizontally through black solid line, then vertically through a violet solid line to point $(0,\theta_1^*)$), to get the upper bound for $\theta_{12}.$ For the lower bound, since $\theta_1+\theta_2-1\geq 0$ if and only if $\theta_2\geq 1-\theta_1,$ then if the chosen point $(\theta_1^*,\theta_2^*)$ is above the solid gray line $\theta_2=1-\theta_1,$ we may build a solid black line $\theta_2=-\theta_1+(\theta_1^*+\theta_2^*)$ that includes the point $(\theta_1^*,\theta_2^*)$ and that is parallel to the solid gray line $\theta_2=1-\theta_1,$ so that $\theta_2=1$ if and only if $\theta_1=\theta_1^*+\theta_2^*-1,$ which is the corresponding lower bound for $\theta_{12},$ and so the point $(\theta_1^*+\theta_2^*-1,1)$ is projected vertically down to $(0,\theta_1^*+\theta_2^*-1).$ So the red thick line represents the interval $[\,\theta_1^*+\theta_2^*-1\,,\,\theta_1^*\,]$ of admissible values for $\theta_{12}$ in this graphical example, given chosen values $\theta_1=\theta_1^*$ and $\theta_2=\theta_2^*.$
    \item[b)] Figure \ref{fig:B2} Right: In this case what is chosen is a specific value $\theta_{12}=\theta_{12}^*$ and now will characterize the set of all points $(\theta_1,\theta_2)$ that are compatible with such fixed value $\theta_{12}^*$ according to (\ref{FH2d}), that is to solve for $\theta_1$ and $\theta_2$ these two inequalities:
    \begin{eqnarray*}
        \max\{\theta_1+\theta_2-1,0\} &\leq& \theta_{12}^*\,,  \\ 
        \min\{\theta_1,\theta_2\} &\geq& \theta_{12}^*\,, 
    \end{eqnarray*}
    which in turn are equivalent to the following inequalities:
    \begin{eqnarray*}
        \theta_2 &\leq& \min\{1, -\theta_1+\theta_{12}^*+1\}\,, \\ 
        \theta_1 &\geq& \theta_{12}^* \,\,\text{ and }\,\, \theta_2 \geq \theta_{12}^*\,,
    \end{eqnarray*}
    where these last two inequalities represent the red triangle region of admissible values for $(\theta_1,\theta_2)$ given a fixed value $\theta_{12}=\theta_{12}^*.$
\end{itemize}

\section{Trivariate Bernoulli}\label{sec:trivariate}

Let $(X_1,X_2,X_3)$ be a vector of Bernoulli random variables with (univariate marginal) parameters $1-\theta_r$ where $0<\theta_r<1$ as explained in the previous section, but now $r\in\{1,2,3\}.$ According to (\ref{sklar}) the domain of the underlying trivariate subcopula is the set $\mbox{Dom}\,S^{(3)}=\{0,\theta_1,1\}\times\{0,\theta_2,1\}\times\{0,\theta_3,1\}$ and:
\begin{equation}\label{subcop3d}
	S^{(3)}(u,v,w) = \begin{cases}
		\,1 & \mbox{ if } (u,v,w)=(1,1,1), \\
		\,\theta_1 & \mbox{ if } (u,v,w)=(\theta_1,1,1), \\
		\,\theta_2 & \mbox{ if } (u,v,w)=(1,\theta_2,1), \\
		\,\theta_3 & \mbox{ if } (u,v,w)=(1,1,\theta_3), \\
		\,\theta_{12} & \mbox{ if } (u,v,w)=(\theta_1,\theta_2,1), \\
		\,\theta_{13} & \mbox{ if } (u,v,w)=(\theta_1,1,\theta_3), \\
		\,\theta_{23} & \mbox{ if } (u,v,w)=(1,\theta_2,\theta_3), \\
		\,\theta_{123} & \mbox{ if } (u,v,w)=(\theta_1,\theta_2,\theta_3), \\
		\,0 & \mbox{ elsewhere.}
	\end{cases}
\end{equation}
Recall from Section \ref{sec:subcopula} that for a trivariate subcopula $S^{(3)}$ there are $\binom{3}{2}=3$ bivariate marginal subcopulas, which in this case have bivariate dependence parameters $\theta_{12},$ $\theta_{13},$ and $\theta_{23},$ respectively, plus a trivariate dependence parameter $\theta_{123}.$ The bivariate dependence parameters must satisfy (\ref{FH2d}), that is:
\begin{equation*}%\label{FH2dbis}
	\max\{\theta_r+\theta_t-1,0\} \,\leq\, \theta_{rt} \,\leq\, \min\{\theta_r,\theta_t\}\,, \quad 1\leq r<t\leq 3\,.
\end{equation*}
The trivariate dependence parameter $\theta_{123}$ must satisfy the Fr\'echet-Hoeffding bounds (\ref{FH}), which in this case translates into:
\begin{equation}\label{FH3d}
	\max\{\theta_1+\theta_2+\theta_3-2,0\} \,\leq\, \theta_{123} \,\leq\, \min\{\theta_1,\theta_2,\theta_3\}\,,
\end{equation}
but it will be discussed later that (\ref{FH3d}) is a necessary but could not be a sufficient condition for an admissible value for $\theta_{123},$ due to a compatibility problem, similarly as mentioned in the bivariate case. Since $\mbox{Ran}\,(X_1,X_2,X_3)=\{(i,j,k):i,j,k\in\{0,1\}\}$ let:
\begin{equation*}%\label{pijk}
	p_{ijk} := \mathbb{P}(X_1=i,X_2=j,X_3=k)\,,\qquad i,j,k\in\{0,1\},
\end{equation*}

\smallskip

\noindent where necessarily $0\leq p_{ijk}\leq 1$ and $\sum_{i,j,k}p_{i,j,k} = 1,$ and therefore only seven of the $p_{ijk}$ need to be specified, since the eighth is just $1$ minus the sum of the other seven. Applying (\ref{sklar}) and (\ref{subcop3d}) to calculate each $p_{ijk}$ in terms of the joint distribution function $F_{123}(X_1\leq i, X_2\leq j, X_3\leq k):$
\begin{eqnarray*}%\label{Bernoulli3dcdf}
	p_{ijk} &=& F_{123}(i,j,k) - F_{123}(i,j,k-1) - F_{123}(i,j-1,k) \ldots \nonumber \\
	&{ }& - F_{123}(i-1,j,k) + F_{123}(i,j-1,k-1) + F_{123}(i-1,j,k-1) \ldots \nonumber \\
	&{ }& + F_{123}(i-1,j-1,k) - F_{123}(i-1,j-1,k-1)\,, \nonumber \\
	&=& S^{(3)}\left(F_1(i),F_2(j),F_3(k)\right) - S^{(3)}\left(F_1(i),F_2(j),F_3(k-1)\right) \ldots \\
	&{ }& - S^{(3)}\left(F_1(i),F_2(j-1),F_3(k)\right) - S^{(3)}\left(F_1(i-1),F_2(j),F_3(k)\right) \ldots \nonumber\\
	&{ }& + S^{(3)}\left(F_1(i),F_2(j-1),F_3(k-1)\right) + S^{(3)}\left(F_1(i-1),F_2(j),F_3(k-1)\right) \ldots \nonumber \\
	&{ }& + S^{(3)}\left(F_1(i-1),F_2(j-1),F_3(k)\right) - S^{(3)}\left(F_1(i-1),F_2(j-1),F_3(k-1)\right)\,, \nonumber
\end{eqnarray*}
and therefore:
\begin{eqnarray}\label{Bernoulli3Dpdf}
	p_{000} &=& \theta_{123}\,, \label{p000} \\ 
	p_{001} &=& \theta_{12} - \theta_{123}\,,\label{p001}  \\
	p_{010} &=& \theta_{13} - \theta_{123}\,,\label{p010}  \\
	p_{100} &=& \theta_{23} - \theta_{123}\,,\label{p100}  \\
	p_{011} &=& \theta_1 - \theta_{12} - \theta_{13} + \theta_{123}\,,\label{p011}\\
	p_{101} &=& \theta_2 - \theta_{12} - \theta_{23} + \theta_{123}\,,\label{p101}\\
	p_{110} &=& \theta_3 - \theta_{13} - \theta_{23} + \theta_{123}\,,\label{p110}\\
	p_{111} &=& 1-\theta_1-\theta_2-\theta_3+\theta_{12}+\theta_{13}+\theta_{23}-\theta_{123}\,,\label{p111}
\end{eqnarray}
where (\ref{p111}) is equal to $1$ minus the sum of (\ref{p000}) through (\ref{p110}). From (\ref{p001}), (\ref{p010}), (\ref{p100}), and (\ref{FH3d}), it is clear that:
\begin{equation*}
	\theta_{123}\,\leq\,\min\{\theta_{12},\theta_{13},\theta_{23}\}\,\leq\,\min\{\theta_1,\theta_2,\theta_3\}
\end{equation*}
a sharper upper bound for $\theta_{123}$ than the Fr\'echet-Hoeffding bounds (\ref{FH3d}), as a consequence of the compatibility problem for the Fr\'echet class $\mathcal{F}\{F_{12},F_{13},F_{23}\},$ but all the inequalities that arise from (\ref{p000}) to (\ref{p111}) have to be analyzed under the restrictions $0\leq p_{ijk}\leq 1$ to solve the full compatibility problem of all the parameters. It is straightforward to obtain expressions for the $\theta$ parameters in terms of the $p_{ijk}$ probabilities:
\begin{eqnarray}
	\theta_{123} &=& p_{000} = \mathbb{P}(X_1=0,X_2=0,X_3=0)\,,\label{th123} \\
	\theta_{12} &=& p_{000} + p_{001} = \mathbb{P}(X_1=0,X_2=0)\,,\label{th12} \\
	\theta_{13} &=& p_{000} + p_{010} = \mathbb{P}(X_1=0,X_3=0)\,,\label{th13} \\
	\theta_{23} &=& p_{000} + p_{100} = \mathbb{P}(X_2=0,X_3=0)\,,\label{th23} \\
	\theta_1 &=& p_{000} + p_{001} + p_{010} + p_{011} = \mathbb{P}(X_1=0)\,,\label{th1} \\
	\theta_2 &=& p_{000} + p_{001} + p_{100} + p_{101} = \mathbb{P}(X_2=0)\,,\label{th2} \\
	\theta_3 &=& p_{000} + p_{010} + p_{100} + p_{110} = \mathbb{P}(X_3=0)\,.\label{th3}
\end{eqnarray}
We may follow the order in (\ref{th123}) through (\ref{th3}) to specify all the parameters such that the full compatibility of the Fr\'echet class $\mathcal{F}\{F_{12},F_{13},F_{23},F_1,F_2,F_3\}$ is guaranteed:

\bigskip

\noindent \textbf{\underline{Parameter Selection Procedure 1}}
\begin{itemize}
	\item[]\underline{Step 1}: Choose $0\leq p_{000}\leq 1$ and set the trivariate dependence parameter $\theta_{123}=p_{000}\,.$
	\item[]\underline{Step 2}: Choose $0\leq p_{001}\leq 1,$ $0\leq p_{010}\leq 1,$ and $0\leq p_{100}\leq 1,$ such that $p_{000}+p_{001}+p_{010}+p_{100}\leq 1,$ and calculate the bivariate dependence parameters $\theta_{rt}$ according to (\ref{th12}), (\ref{th13}), and (\ref{th23}).
	\item[]\underline{Step 3}: Choose $0\leq p_{011}\leq 1,$ $0\leq p_{101}\leq 1,$ and $0\leq p_{110}\leq 1,$ such that $p_{000}+p_{001}+p_{010}+p_{100}+p_{011}+p_{101}+p_{110}\leq 1,$ and calculate the univariate (marginal) parameters $\theta_r$ according to (\ref{th1}), (\ref{th2}), and (\ref{th3}).
	\item[]\underline{Step 4}: Calculate $p_{111}=1-p_{000}-p_{001}-p_{010}-p_{011}-p_{100}-p_{101}-p_{110}\,.$
\end{itemize}

Finally, we may apply (\ref{depmeas2d}) to calculate the bivariate dependence measures $\mu_{rt}=\mu(X_r,X_t)$ for $1\leq r<t\leq 3,$ and following the same idea:
\begin{equation}\label{depmeas3d}
	\mu_{123} =  \begin{cases}
		\frac{\theta_{123}-\theta_1\theta_2\theta_3}{\min\{\theta_1,\theta_2,\theta_3\}-\theta_1\theta_2\theta_3} & \mbox{ if } \theta_{123}\geq \theta_1\theta_2\theta_3 \,, \\
		{ } &     { } \\
		\frac{\theta_{123}-\theta_1\theta_2\theta_3}{\theta_1\theta_2\theta_3 - \max\{\theta_1+\theta_2+\theta_3-2,0\}} & \mbox{ if } \theta_{123} <  \theta_1\theta_2\theta_3 \,.
	\end{cases}
\end{equation}
where from (\ref{FH3d}) $-1\leq\mu_{123}\leq +1,$ with $\mu_{123}=0$ if and only if $\theta_{123}=\theta_{1}\theta_{2}\theta_{3}.$ This characterizes de absence of three-way interaction, not independence: the three random variables are jointly independent if and only if $\mu_{12}=\mu_{13}=\mu_{23}=\mu_{123}=0$ all hold. 
%% where from (\ref{FH3d}) $-1\leq\mu_{123}\leq +1,$ and clearly $\mu_{123}=0$ if and only if the three random variables are independent.

\bigskip

\textbf{\textit{Example 1.}} Consider $X_1,X_2,X_3$ identically distributed Bernoulli random variables with parameter $1-\theta,$ where $0<\theta<1,$ that are pairwise independent but not necessarily jointly independent. We will obtain all the compatible values for the trivariate dependence parameter $\theta_{123},$ and the resulting trivariate probability mass function (pmf). Combining the inequalities $0\leq p_{ijk}\leq 1$ (\ref{p000}) through (\ref{p111}) with $\theta_r=\theta$ and $\theta_{rt}=\theta^2$ (pairwise independence) we obtain:
\begin{equation*}%\label{eq:ex1a}
	\max\{0,2\theta^2-\theta\} \,\leq\, \theta_{123} \,\leq\, \min\{\theta^2,1+\theta-2\theta^2,1-3\theta+3\theta^2\}
\end{equation*}
which is equivalent to:
\begin{equation}\label{eq:ex1b}
	\theta_{123} \in \begin{cases}
		\,\left[0\,,\,\theta^2\right] & \text{ if } \theta\leq\frac{1}{2} \\ 
		\,\left[2\theta^2-\theta\,,\,1-3\theta+3\theta^2\right] & \text{ if } \frac{1}{2}<\theta\leq\frac{4}{5} \\
		\,\left[2\theta^2-\theta\,,\,1+\theta-2\theta^2\right] & \text{ if } \frac{4}{5}<\theta\leq\frac{1+\sqrt{5}}{4} 
	\end{cases}
\end{equation}

\begin{figure}
	\begin{center} 
		\includegraphics[width=8cm, keepaspectratio]{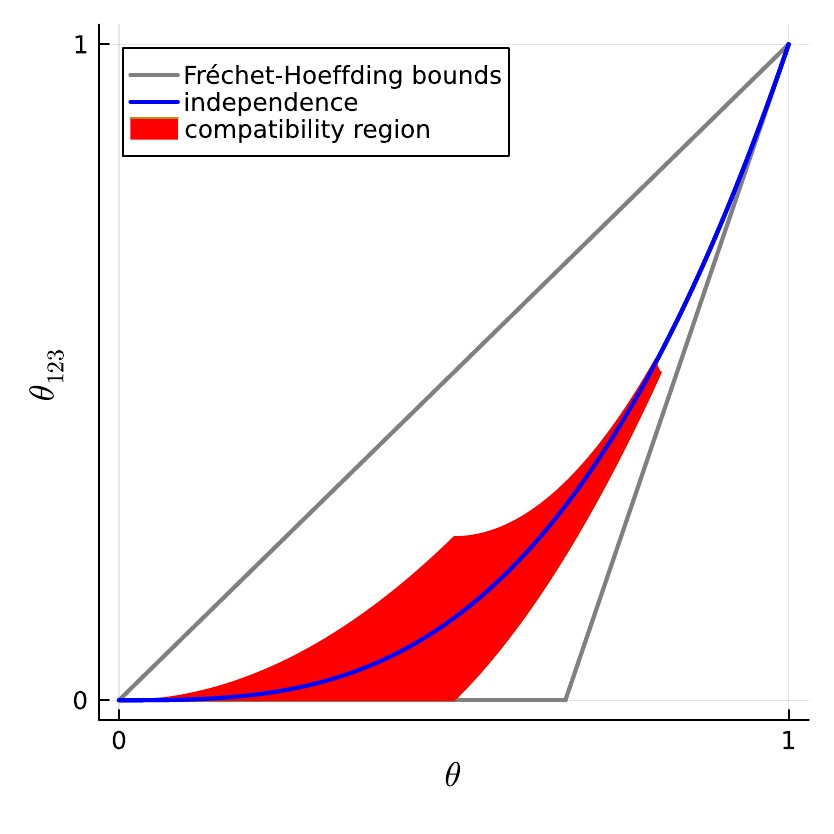}
	\end{center}
	\caption{Compatible values $(\theta,\theta_{123})$ (red region) in Example 1 for a Trivariate Bernoulli, with identical univariate marginals with parameter $1-\theta,$ pairwise independent, with trivariate dependence parameter $\theta_{123}.$}
	\label{fig:ex2}
\end{figure}

In Figure \ref{fig:ex2} it is depicted the set of all compatibility values $(\theta,\theta_{123}),$ and in Table \ref{tab:ex1} the resulting pmf in the particular case when $\theta=1/2$ and the special cases of extreme dependence ($\theta_{123}\in\{0,1/4\}$) and independence ($\theta_{123}=1/8$).

\begin{table}
	\begin{tabular}{|c|c|c|c|c|} \hline
		{       } & {            } & $\mu_{123}=-1$   & $\mu_{123}=0$      & $\mu_{123}=+1/3$ \\
		$(i,j,k)$ & $p_{ijk}$ & $\theta_{123}=0$ & $\theta_{123}=1/8$ & $\theta_{123}=1/4$ \\ \hline\hline  
		$(0,0,0)$ & $\theta_{123}$ & $0$ & $1/8$ & $1/4$ \\
		$(0,0,1)$ & $\theta^2-\theta_{123}$ & $1/4$ & $1/8$ & $0$ \\ 
		$(0,1,0)$ & $\theta^2-\theta_{123}$ & $1/4$ & $1/8$ & $0$ \\ 
		$(1,0,0)$ & $\theta^2-\theta_{123}$ & $1/4$ & $1/8$ & $0$ \\
		$(0,1,1)$ & $\theta-2\theta^2+\theta_{123}$ & $0$ & $1/8$ & $1/4$ \\
		$(1,0,1)$ & $\theta-2\theta^2+\theta_{123}$ & $0$ & $1/8$ & $1/4$ \\
		$(1,1,0)$ & $\theta-2\theta^2+\theta_{123}$ & $0$ & $1/8$ & $1/4$ \\
		$(1,1,1)$ & $1-3\theta+3\theta^2-\theta_{123}$ & $1/4$ & $1/8$ & $0$ \\ \hline 
	\end{tabular}
	\caption{Probability mass function of a trivariate Bernoulli in Example 1, with identical univariate marginals with parameter $1-\theta=1/2,$ pairwise independent, and with trivariate parameter $\theta_{123}\in\{0,1/8,1/4\}$ corresponding to the lowest compatible value, independence, and the highest compatible value, respectively.} 
	\label{tab:ex1}
\end{table}

\bigskip

\textbf{\textit{Example 2.}} Algorithm 1 will be applied to obtain non independent $X_1,X_2,X_3$ but identically distributed Bernoulli random variables with parameter $1-\theta,$ where $0<\theta<1,$ with pairwise conditional independence, but not pairwise independent. For the sake of simplicity we will assume that the bivariate dependence parameter is the same for all pairs, say $\beta:=\theta_{12}=\theta_{13}=\theta_{23}$ and let $\tau:=\theta_{123}$ be the trivariate dependence parameter.
\begin{itemize}
	\item[]\underline{Step 1}: Choose $p_{000}\in[0,1]$ and set the trivariate dependence parameter $\tau=p_{000}.$ 
	\item[]\underline{Step 2}: Choose $p_{001},p_{010},p_{100}\in[0,1]$ such that:
	\begin{equation*}\label{ex2:eq1}
		p_{000}+p_{001}+p_{010}+p_{100}\leq 1
	\end{equation*}
	but from (\ref{p001}), (\ref{p010}), and (\ref{p100}), we get $p_{001}=\beta-\tau=p_{010}=p_{100}$ and therefore $\beta\geq\tau.$ From (\ref{ex2:eq1}) $p_{001}+p_{010}+p_{100}\leq 1-\tau$ implies $\beta\leq(1+2\tau)/3,$ so combining altogether:
	\begin{equation}\label{ex2:eq2}
		\tau\,\leq\,\beta\,\leq\,\frac{1+2\tau}{3}
	\end{equation}
	\item[]\underline{Step 3}: Choose $p_{011},p_{101},p_{110}\in[0,1]$ such that:
	\begin{equation}\label{ex2:eq3}
		p_{000}+p_{001}+p_{010}+p_{100}+p_{011}+p_{101}+p_{110}\leq 1
	\end{equation}
	but from (\ref{p011}), (\ref{p101}), and (\ref{p110}), we get $p_{011}=\theta-2\beta+\tau=p_{101}=p_{110}$ and therefore from (\ref{ex2:eq3}) we obtain:
	\begin{equation}\label{ex2:eq4}
		\theta\,\leq\,\frac{1-\tau}{3}+\beta 
	\end{equation}
	\item[]\underline{Step 4}: Calculate $p_{111}=1-\tau-3(\beta-\tau)-3(\theta-2\beta+\tau)=1-\tau+3(\beta-\theta).$
\end{itemize}

In summary, once specified the trivariate dependence parameter $0\leq\tau\leq 1$ then we have to choose the bivariate dependence parameter $\beta$ according to (\ref{ex2:eq2}) and the univariate parameter $\theta$ according to (\ref{ex2:eq4}), for compatibility. If in addition we want pairwise conditional independence, say for example $X_1$ and $X_2$ conditionally independent given the value of $X_3:$
\begin{multline*}
	X_1\perp X_2\,|\,X_3=k \,\,\text{ where } k\in\{0,1\} \quad \Leftrightarrow \\
	\begin{cases}
		\mathbb{P}(X_1=0,X_2=0\,|\,X_3=0) = \mathbb{P}(X_1=0\,|\,X_3=0)\mathbb{P}(X_2=0\,|\,X_3=0) \\
		\mathbb{P}(X_1=0,X_2=0\,|\,X_3=1) = \mathbb{P}(X_1=0\,|\,X_3=1)\mathbb{P}(X_2=0\,|\,X_3=1)
	\end{cases}
\end{multline*}
\begin{eqnarray}
	X_1\perp X_2\,|\,X_3=k &\Leftrightarrow& \begin{cases}
		\frac{p_{000}}{\theta} = \frac{p_{000}+p_{010}}{\theta}\cdot\frac{p_{000}+p_{100}}{\theta} \\
		\frac{p_{001}}{1-\theta} = \frac{p_{001}+p_{011}}{1-\theta}\cdot\frac{p_{001}+p_{101}}{1-\theta}
	\end{cases} \nonumber \\
	&\Leftrightarrow& \begin{cases}
		\theta\tau = \beta^2 \\
		(1-\theta)(\beta-\theta) = (\theta-\beta)^2
	\end{cases} \nonumber \\
	&\Leftrightarrow& \tau^2(\beta-\tau) = \beta^3(\beta-\tau) \label{ex2:eq5}			 
\end{eqnarray}
where (\ref{ex2:eq5}) is always satisfied whenever $\beta=\tau$ which in turn would imply that $\theta=\beta$ and since $\beta\neq\theta^2$ nor $\tau=\theta^3$ for $0<\theta<1$ then the condition $\theta=\beta=\tau$ implies pairwise dependence and trivariate dependence but with pairwise conditional independence. In fact, in terms of the probabilities $p_{ijk}$ this implies $p_{000}=\theta,$ $p_{111}=1-\theta$ and zero probability for the remaining six cases.

\section{Multivariate Bernoulli}

Let $(X_1,\ldots,X_n)$ be a vector of $n\geq 2$ Bernoulli random variables with (univariate marginal) parameters $1-\theta_r$ where $0<\theta_r<1,$ $r\in\{1,\ldots,n\}.$ According to (\ref{sklar}) the domain of the underlying multivariate subcopula is the set $\operatorname{Dom}\,S^{(n)} = \{0,\theta_1,1\} \,\times\, \cdots\,\times\, \{0,\theta_n,1\}$ and:
\begin{equation*}%\label{subcopnd}
	S^{(n)}(u_1,\ldots,u_n) = \begin{cases}
		\,1 & \mbox{ if } u_r=1 \mbox{ for all } r\in\{1,\ldots,n\}, \\
		\,\theta_s & \mbox{ if } u_s=\theta_s \mbox{ and } u_r=1 \mbox{ for all } r\neq s, \\
		\,\theta_{rs} & \mbox{ if } r<s, u_r=\theta_r, u_s=\theta_s, \mbox{ and } u_t=1 \mbox{ for all } t \neq r,s, \\
		\,\theta_{rst} & \mbox{ if } r<s<t, u_r=\theta_r, u_s=\theta_s, u_t=\theta_t, u_{\ell}=1, \ell\neq r,s,t, \\
		\quad\vdots & \qquad\vdots \\
		\,\theta_{12\cdots n} & \mbox{ if } u_1=\theta_1,\ldots,u_n=\theta_n, \\
		\,0 & \mbox{ elsewhere.}
	\end{cases}
\end{equation*}
with a total of $2^{n}-1$ parameters. For any ordered subset of $\{1,\ldots,n\},$ say $A=\{a_1,\ldots,a_m\}$ where $m\leq n$ and if $r<s$ then $a_r<a_s,$ the $m$-variate dependence parameter $\theta_{\! A}$ must satisfy the Fr\'echet-Hoeffding bounds (\ref{FH}), which in this case translates into:
\begin{equation}\label{FHnd}
	\max\left\{\sum_{a\,\in\,A}\theta_a -m + 1\,,\,0\right\} \,\leq\, \theta_A \,\leq\, \min\{\theta_a:a\in A\}\,.
\end{equation}
Since $\mbox{Ran}\,(X_1,\ldots,X_n)=\{0,1\}^n$ let:
\begin{equation}\label{pi1in}
	p_{i_1\ldots i_n} := \mathbb{P}(X_1=i_1,\ldots,X_n=i_n)\,,\qquad i_r\in\{0,1\},
\end{equation}
where necessarily $0\leq p_{i_1\ldots i_n}\leq 1$ and $\sum\cdots\sum p_{i_1\ldots i_n} = 1,$ and therefore just $2^{n}-1$ of the $p_{i_1\ldots i_n}$ need to be specified, since the $2^n$-th is just $1$ minus the sum of the other $2^{n}-1.$ Then expressions for the $\theta_{\! A}$ parameters in terms of (\ref{pi1in}) probabilities are obtained by:
\begin{equation}\label{thetaA} 
	\theta_{\! A} \,=\, \!\!\!\!\!\!\!\!\!\!\!\!\sum_{\qquad\qquad i_r\,\notin\,A}\!\!\!\!\!\!\!\!\!\!\!\!\cdots\sum p_{i_1\ldots i_n}
\end{equation}
which can be specified according to the following algorithm to guarantee full compatibility of the dependence parameters, with the conventional notation that if $D$ is a set of numbers then $\sum D$ is equivalent to $\sum_{d\in D} d\,:$

\bigskip 

\noindent \textbf{\underline{Parameter Selection Procedure 2}}
\begin{itemize}
	\item[]\underline{Step 1}: Choose $0\leq p_{0\cdots 0}\leq 1,$ and define the singleton set $B_0:=\{p_{0\cdots 0}\}.$ Set the $n$-variate dependence parameter $\theta_{1\cdots n}=p_{0\cdots 0}.$
	\item[]\underline{Step 2}: Choose $B_1:=\{0\leq p_{i_1\ldots i_n}\leq 1:\sum i_r=1\}$ such that $\sum B_0\cup B_1\leq 1,$ and set the $(n-1)$-variate parameters $\theta_{\!A}$ according to (\ref{thetaA}).
	\item[]\underline{Step 3}: Choose $B_2:=\{0\leq p_{i_1\ldots i_n}\leq 1:\sum i_r=2\}$ such that $\sum B_0\cup B_1\cup B_2\leq 1,$ and set the $(n-2)$-variate parameters $\theta_{\!A}$ according to (\ref{thetaA}). \newline 
	\vdots 
	\item[]\underline{Step $k+1$}: Choose $B_k:=\{0\leq p_{i_1\ldots i_n}\leq 1:\sum i_r=k\}$ such that $\sum\bigcup_{t\,=\,0}^{k}B_t\leq 1,$ and set the $(n-k)$-variate parameters $\theta_{\!A}$ according to (\ref{thetaA}). \newline 
	\vdots 
	\item[]\underline{Step $n+1$}: Calculate $p_{1\cdots 1} = 1 - \sum\bigcup_{t\,=\,0}^{n-1}B_t\,.$
\end{itemize}

Finally, we may calculate $m$-variate dependence measures for $2\leq m\leq n$ as follows:
\begin{equation}\label{depmeasnd}
	\mu_A =  \begin{cases}
		\frac{\theta_{\! A} - \prod_{r\in A}\theta_r}{\min\{\theta_a:a\in A\} - \prod_{r\in A}\theta_r} & \mbox{ if } \theta_{\! A} \geq \prod_{r\in A}\theta_r \,, \\
		{ } &     { } \\
		\frac{\theta_{\! A} - \prod_{r\in A}\theta_r}{\prod_{r\in A}\theta_r - \max\{\sum_{a\in A}\theta_a - m + 1,0\}} & \mbox{ if } \theta_{\! A} <  \prod_{r\in A}\theta_r\,.
	\end{cases}
\end{equation}
where from (\ref{FHnd}) $-1\leq\mu_{A}\leq +1,$ with $\mu_A=0$ if and only if $\theta_A=\prod_{r\in A}\theta_r$, i.e. no $|A|$-th order interaction is present. The full vector $(X_1,\ldots,X_n)$ is mutually independent if and only if $\mu_A=0$ for every $A$ with $|A|\geq 2.$
%% where from (\ref{FHnd}) $-1\leq\mu_{A}\leq +1,$ and clearly $\mu_{A}=0$ if and only if the random variables indexed by $A$ are independent.

\subsection{Bayesian inference}

%%%%%% In response to Reviewers 1 and 2
Under the Bayesian paradigm of statistics, given an observed random sample $\mathbf{y}=(y_1,\ldots,y_m)$ from a probability density model $f_Y(y\,|\,\theta),$ where $\theta\in\Theta$ is a vector of unknown parameters, any statistical inference about $\theta$ is performed through a posterior distribution for $\theta$ given the data $\mathbf{y}$ applying Bayes's rule, that is:
\begin{equation*}
	\pi(\theta\,|\,\mathbf{y}) = \frac{f(\mathbf{y}\,|\,\theta)\pi(\theta)}{\int_{\,\,\,\Theta}\!\!\!\!\!\!\cdots\!\int f(\mathbf{y}\,|\,\tilde{\theta})\pi(\tilde{\theta})\,d\tilde{\theta}}
\end{equation*}
where $\pi(\theta)$ is a prior distribution, and where:
\begin{equation*}
	f(\mathbf{y}\,|\,\theta) \,=\, \prod_{k\,=\,1}^m f_Y(y_k\,|\,\theta) \,=\, L(\theta\,|\,\mathbf{y})
\end{equation*}
is also known as the likelihood function. Whenever possible, it is quite convenient to choose as a prior distribution a conjugate family, such that the posterior belongs to that same family, and therefore it is straightforward to identify, since:
\begin{equation*}
	\pi(\theta\,|\,\mathbf{y}) \,\propto\,L(\theta\,|\,\mathbf{y})\pi(\theta)\,.
\end{equation*}
%%%%

\medskip

If $\{\mathbf{x}_1,\ldots,\mathbf{x}_m\}$ is a size $m$ observed random sample from a Multivariate Bernoulli random vector $\mathbf{X}=(X_1,\ldots,X_n)$ where $\mathbf{x}_j=(x_{j,1},\ldots,x_{j,n})$ for $j\in\{1,\ldots,m\},$ we can make statistical inferences about the $2^{n}$ parameters $p_{i_1\ldots i_n}.$ For a more simplified notation, following \cite{bib1}, since $(i_1,\ldots,i_n)\in\{0,1\}^n$ we may consider it as a $n$-positions binary number and apply the one-to-one mapping conversion to a positive integer. 
%\begin{equation}\label{binary}
%r \equiv r(i_1,\ldots,i_n) = \sum_{k\,=\,1}^{n}2^{n-k}i_{k}\,.
%\end{equation}
For $r\in\{1,\ldots,2^n-1\}$ let $B(r-1)=[i_1(r),\ldots,i_n(r)]$ be the $n$-positions binary representation of $r-1$, so that:
\begin{equation*}%\label{binary1}
	p_r = \mathbb{P}[X_1=i_1(r),\ldots,X_n=i_n(r)]
\end{equation*}
and:
\begin{equation*}%\label{binary2}
	p_{2^n} = \mathbb{P}[X_1=1,X_2=1,\ldots,X_n=1] = 1 - \sum_{r\,=\,1}^{2^n-1}p_r\,.
\end{equation*}
Then the likelihood function associated with the unknown parameters $p_1,\ldots,p_{2^n-1}$ is given by:
\begin{equation}\label{likelihood1}
	\mathbb{L}(p_1,\ldots,p_{2^n-1}\,|\,\mathbf{x}_1,\ldots,\mathbf{x}_m) = \left(\prod_{r\,=\,1}^{2^n-1}p_r^{\,n_r}\right)\left(1-\sum_{r\,=\,1}^{2^n-1}p_r\right)^{m-\sum_{r\,=\,1}^{2^n-1}n_r}
\end{equation}
where:
\begin{equation*}%\label{likelihood2}
	n_r = \sum_{j\,=\,1}^m \mathbf{1}_{\{\mathbf{x}_j\,=\,B(r-1)\}}\,.
\end{equation*}
Clearly (\ref{likelihood1}) has as conjugate prior distribution a $2^n-1$ dimensional Dirichlet (Di) distribution with $2^n$ hyperparameters $\alpha_r>0,$ see for example \cite{bib16}, and therefore the posterior distribution is given by:
\begin{equation}\label{posterior}
	\pi(\mathbf{p}\,|\,\mathbf{x}) \sim \text{Di}_{\,2^n-1}(\alpha_1+n_1,\ldots,\alpha_{2^n-1}+n_{2^n-1},\alpha_{2^n}+m - \sum_{r\,=\,1}^{2^n-1}n_r)
\end{equation}
where $\mathbf{p}=(p_1,\ldots,p_{2^n-1})$ and $\mathbf{x}=(\mathbf{x}_1,\ldots,\mathbf{x}_m).$ With (\ref{posterior}) we can also make posterior inferences for the $\theta_{\!A}$ parameters through (\ref{thetaA}), and for the $\mu_{A}$ parameters through (\ref{depmeasnd}).

\section{Computational implementation}\label{JuliaMBD}

The Julia programming language \cite{bib6} has been used for the implementation of all the previous results, and the code is available in a public repository \cite{bib17} to ensure full reproducibility. Julia offers significant advantages for high-performance numerical computing, especially for high-dimensional matrix operations and intensive statistical computations. Its speed and ability to handle large datasets efficiently make it particularly suitable for the implementation of the proposed methods. Additionally, Julia's syntax is close to mathematical notation, facilitating the translation of theoretical concepts into computational code, see for example \cite{bib18}. The main functions implemented:

\begin{itemize}
	\item \texttt{MBerDep} Calculates dependence parameters $\theta_A$ and measures $\mu_A$ given the parameters $p_1,\ldots,p_{2^n}$ of a $n$-dimensional multivariate Bernoulli distribution.
	\item \texttt{MBerMargin} Calculates dependence parameters and measures for the $m$-dimensional marginal distribution of a $n$-dimensional multivariate Bernoulli distribution, where $1\leq m\leq n.$
	\item \texttt{MBerCond} Calculates conditional probabilities, dependence parameters and measures from a $n$-dimensional
	multivariate Bernoulli distribution, given the values for a subset of random variables.
	\item \texttt{MBerSim} Simulates a random sample from a $n$-dimensional multivariate Bernoulli distribution.
	\item \texttt{MBerBayes} Posterior Dirichlet model for the parameters of a $n$-dimensional multivariate Bernoulli distribution 
	given an observed random sample.
	\item \texttt{MBerInf} Posterior point and interval estimation for the probability parameters, and the dependence parameters and measures, given an observed sample from a multivariate Bernoulli distribution, and a prior value. For dimension $10$ or higher calculations may take more than an hour. In such case you may consider using \texttt{MBerEst} just for point estimations.
\end{itemize}

%%%% In response to Reviewer 1 we have now 3 subsubsections:
%%%% 5.2.1 A theoretical example
%%%% 5.2.2 A simulation example 
%%%% 5.2.3 A real data example

\subsection{A theoretical example}\label{ex:theoretical}

In Example 1 from Section \ref{sec:trivariate}, %if we choose $\theta=0.6$ then according to (\ref{eq:ex1b}) we must choose $\theta_{123}\in[0.12,0.28]$ where $\theta^3=0.216$ belongs to such interval and would represent the case of joint independence. Let's use $\theta_{123}=0.15$ and 
%% In response to Editor 3)
the parameter $\theta$ fixes the common marginals $\theta_r=\mathbb{P}(X_r=0)$ and, under the restriction $\theta_{12}=\theta_{13}=\theta_{23}=\theta^2,$ the remaining free parameter $\theta_{123}$ must satisfy the compatibility bounds in (\ref{eq:ex1b}). If we choose $\theta=0.6\in\,]0.5,0.8]$ we obtain $\theta_{123}\in[0.12,0.28],$ and because $\theta^3=0.216$ (the joint independence value) lies strictly inside this interval, selecting $\theta_{123}=0.15\neq\theta^3$ creates nontrivial three-way dependence without breaking pairwise independence. Applying formulas (\ref{p000}) through (\ref{p111}) we may obtain the $p_{ijk}$ joint probabilities $p_{000}=0.15,$ $p_{001}=0.21,$ $p_{010}=0.21,$ $p_{011}=0.03,$ $p_{100}=0.21,$ $p_{101}=0.03,$ $p_{110}=0.03,$ and $p_{111}=0.13.$ The Julia code to generate the results from Table \ref{tab:ex1bis} is in Appendix \ref{secA1}, which basically makes use of the \texttt{MBerDep} function described at the beginning of Section \ref{JuliaMBD}.

\begin{table}[h]
	\begin{tabular}{|c|c|c|} \hline
		variables       & dependence parameter & dependence measure \\ \hline\hline  
		$(X_1,X_2,X_3)$ & 0{.}15 & -0{.}3056 \\
		$(X_1,X_2)$     & 0{.}36 & 0{.}0 \\ 
		$(X_1,X_3)$     & 0{.}36 & 0{.}0 \\ 
		$(X_2,X_3)$     & 0{.}36 & 0{.}0 \\ \hline 
	\end{tabular}
	\caption{Dependence parameters and measures corresponding to the provided values in subsection \ref{ex:theoretical}. The Julia code to generate the results is in Appendix \ref{secA1}.} 
	\label{tab:ex1bis}
\end{table}

All pairwise dependencies $\mu_{ij}$ in Table \ref{tab:ex1bis} are equal to zero (as expected) but with a 3-variate negative dependence $\mu_{123}=-0.306$ as expected, since we chose a value $\theta_{123}=0.15$ below the value that would represent independence ($\theta^3=0.216$).

\subsection{A simulation example}\label{ex:simulated}

A trivariate Bernoulli distribution will be defined applying \textit{Parameter Selection Procedure 1} from Section \ref{sec:trivariate} as follows:
\begin{itemize}
	\item[] Step 0: Choose $p_{000} = 0{.}1$
	\item[] Step 1: Choose $(p_{001}, p_{010}, p_{100}) = (0{.}2, 0{.}1, 0{.}05)$
	\item[] Step 2: Choose $(p_{011}, p_{101}, p_{110}) = (0{.}2, 0{.}15, 0{.}1)$
	\item[] Step 3: Calculate $p_{111} = 1 - \sum_{ijk\neq 111} p_{ijk}$
\end{itemize}

The Julia code to define a MBD, simulate data from it, and make statistical inferences is in Appendix \ref{secA2}. A random sample of size $3,000$ is simulated, according to the above provided probabilities, using the \texttt{MBerSim} function described at the beginning of Section \ref{JuliaMBD}, which in turn uses a standard procedure to simulate from a discrete distribution with given mass probabilities. With the simulated data, point and interval estimations are made using the \texttt{MBerInf} function, for the probability parameters (Table \ref{tab:ex3A}), the parameters $\theta_A$ (Table \ref{tab:ex3B}), and the dependence measures $\mu_A$ (Table \ref{tab:ex3C}).

To evaluate the finite-sample performance of the proposed estimation procedure, a Monte Carlo study was conducted and examined the empirical coverage rates of the bayesian probability intervals for the parameters of interest. For each scenario, data were generated under the given parameter values, and 10,000 replications were performed. A 99\% probability interval was constructed in each replication, and the proportion of intervals containing the true parameter value was recorded. Ideally, this proportion should be close to the nominal level of 99\%, and this is the case, since conjugate analysis was applied (Dirichlet distribution in this case), and there are no convergence issues for the same reason, since there was no need of applying MCMC algorithms for the posterior inferences.

\begin{table}[h]
	\begin{tabular}{|c|c|c|c|c|c|} \hline
		{       } & {true value} & Lower & Median & Upper & Coverage \\
		$(i,j,k)$ & $p_{ijk}$ & quantile $0{.}005$ & quantile $0{.}500$ & quantile $0{.}995$ & rate (\%) \\ \hline\hline  
		$(0,0,0)$ & $0{.}10$ & 0.0862 & 0.0996 & 0.1143 & 98.95 \\
		$(0,0,1)$ & $0{.}20$ & 0.1767 & 0.1948 & 0.2138 & 98.98 \\ 
		$(0,1,0)$ & $0{.}10$ & 0.0874 & 0.1010 & 0.1158 & 99.05 \\ 
		$(1,0,0)$ & $0{.}05$ & 0.0333 & 0.0420 & 0.0520 & 99.08 \\
		$(0,1,1)$ & $0{.}20$ & 0.1751 & 0.1935 & 0.2124 & 99.02 \\
		$(1,0,1)$ & $0{.}15$ & 0.1391 & 0.1555 & 0.1733 & 98.92 \\
		$(1,1,0)$ & $0{.}10$ & 0.0905 & 0.1043 & 0.1193 & 99.09 \\
		$(1,1,1)$ & $0{.}10$ & 0.0946 & 0.1086 & 0.1238 & 99.15 \\ \hline 
	\end{tabular}
	\caption{Comparing theoretical probabilities (true values) with point estimates (median) and $99\%$ probability intervals (lower--upper) from simulated observations of the Multivariate Bernoulli Distribution defined in section \ref{ex:simulated}. For the coverage rate 10,000 replications were performed in each case. The Julia code to generate the results is in Appendix \ref{secA2}.} 
	\label{tab:ex3A}
\end{table}

\begin{table}[h]
	\begin{tabular}{|c|c|c|c|c|c|} \hline
		{       }              & {true value} & Lower & Median & Upper & Coverage \\
		$A\subseteq\{1,2,3\}$  & $\theta_{\!A}$   & quantile $0{.}005$ & quantile $0{.}500$ & quantile $0{.}995$ & rate (\%) \\ \hline\hline  
		$\{1,2,3\}$            & $0{.}10$     & 0.0862 & 0.0996 & 0.1143 & 98.95 \\
		$\{1,2\}$              & $0{.}30$     & 0.2734 & 0.2945 & 0.3162 & 99.02 \\ 
		$\{1,3\}$              & $0{.}20$     & 0.1823 & 0.2007 & 0.2199 & 98.97 \\ 
		$\{2,3\}$              & $0{.}15$     & 0.1260 & 0.1417 & 0.1586 & 99.05 \\
		$\{1\}$                & $0{.}60$     & 0.5662 & 0.5892 & 0.6123 & 98.93 \\
		$\{2\}$                & $0{.}50$     & 0.4688 & 0.4923 & 0.5160 & 99.18 \\
		$\{3\}$                & $0{.}35$     & 0.3250 & 0.3472 & 0.3697 & 98.91 \\ \hline 
	\end{tabular}
	\caption{Comparing theoretical parameters $\theta_A$ (true values) with point estimates (median) and $99\%$ probability intervals (lower--upper) from simulated observations of the Multivariate Bernoulli Distribution defined in section \ref{ex:simulated}. For the coverage rate 10,000 replications were performed in each case. The Julia code to generate the results is in Appendix \ref{secA2}.} 
	\label{tab:ex3B}
\end{table}

\begin{table}[h]
	\begin{tabular}{|c|r|r|r|r|c|} \hline
		{       }              & {true value} & Lower & Median & Upper & Coverage \\
		$A\subseteq\{1,2,3\}$  & $\mu_A$      & quantile $0{.}005$ & quantile $0{.}500$ & quantile $0{.}995$ & rate(\%) \\ \hline\hline  
		$\{1,2,3\}$            & -0.0476  & -0.1131 & -0.0104 &  0.0391 & 99.15 \\
		$\{1,2\}$              &  0.0000  & -0.0338 &  0.0223 &  0.0794 & 99.04 \\ 
		$\{1,3\}$              & -0.0476  & -0.0725 & -0.0188 &  0.0505 & 99.06 \\ 
		$\{2,3\}$              & -0.1429  & -0.2349 & -0.1705 & -0.1051 & 99.03 \\ \hline
	\end{tabular}
	\caption{Comparing theoretical dependence measures $\mu_A$ (true values) with point estimates (median) and $99\%$ probability intervals (lower--upper) from simulated observations of the Multivariate Bernoulli Distribution defined in section \ref{ex:simulated}. For the coverage rate 10,000 replications were performed in each case. The Julia code to generate the results is in Appendix \ref{secA2}.} 
	\label{tab:ex3C}
\end{table}

\subsection{A real data example: Bank churn data}

Data related to churn detection in the banking sector has been analyzed by \cite{Vaduva}: 10,000 registries where 4 out of 13 variables under consideration are binary. The original data may be downloaded from \cite{Meshram}, but for the present example the 4 mentioned binary variables are available for download from \cite{bib17} as a \texttt{churnbindata.csv} file, with the following description of the variables:
\begin{enumerate}
    \item \texttt{gender}: 1=male, 0=female.
    \item \texttt{crcard}: customer has credit card (1=yes, 0=no).
    \item \texttt{active}: customer is an active member (1=yes, 0=no).
    \item \texttt{exited}: customer has churned (1=yes, 0=no).
\end{enumerate}

%% Answer to Editor comment 4)
In a retail bank, churn models are often deployed as AI/ML scoring tools that rank customers by the probability of exiting. The present contribution is useful as an interpretable dependence layer on top of (or alongside) such tools: (i) it quantifies not only pairwise associations but also higher-order interaction effects among binary customer attributes; (ii) it produces transparent conditional-probability “rules” (e.g., churn risk given specific combinations of activity, product count, credit card, etc.) that can be audited and explained to non-technical stakeholders; and (iii) it can be used as a diagnostic/sanity-check for black-box models (detecting when a model’s strongest signals correspond to genuine joint-dependence patterns in the data). This supports better targeting of retention actions and better governance (explainability, monitoring, and stability checks).

The Julia code to fit a MBD is in Appendix \ref{secA3}. In this case $n=4$ and therefore there are $2^{n}=16$ probabilities to estimate, $2^{n}-1=15$ parameters $\theta_{\!A}$ to estimate, and $2^{n}-(n+1)=11$ multivariate dependencies $\mu_A$ to estimate (pairwise, three-wise, and four-wise), where the latter are summarized in Table \ref{tab:ex5A}, with point and 95\% probability interval estimations.
\begin{table}[h]
	\begin{tabular}{|c|r|r|r|} \hline
		{       }              & $\mu_A$ lower & $\mu_A$ median & $\mu_A$ upper  \\
		$A\subseteq\{1,2,3\}$  & quantile $0{.}025$ & quantile $0{.}500$ & quantile $0{.}975$ \\ \hline\hline  
        $\{1, 2, 3, 4\}$     & -0.1926 & -0.1254 & -0.0557 \\
        $\{1, 2, 3\}$        & -0.0528 &  0.0022 &  0.0198 \\ 
        $\{1, 2, 4\}$        & -0.0956 & -0.0545 & -0.0131 \\
        $\{1, 3, 4\}$        & -0.1537 & -0.1252 & -0.0966 \\
        $\{2, 3, 4\}$        & -0.1233 & -0.0847 & -0.0457 \\ 
        $\{1, 2\}$           & -0.0230 &  0.0085 &  0.0361 \\ 
        $\{1, 3\}$           &  0.0032 &  0.0240 &  0.0448 \\
        $\{2, 3\}$           & -0.0502 & -0.0189 &  0.0116 \\
        $\{1, 4\}$           & -0.2267 & -0.1914 & -0.1559 \\ 
        $\{2, 4\}$           & -0.0347 & -0.0093 &  0.0373 \\
        $\{3, 4\}$           & -0.3356 & -0.2991 & -0.2623 \\ \hline 
	\end{tabular}
	\caption{Dependencies $\mu_A$ of all orders for the binary variables from bank churn data in \cite{Meshram}. The Julia code to generate the results is in Appendix \ref{secA3}.} 
	\label{tab:ex5A}
\end{table}

The main variable of interest is \texttt{exited} (variable 4) to be predicted by the others. In terms of pairwise dependencies, we may notice that \texttt{crcard} (variable 2) has no relevant dependence with \texttt{exited} since $\mu_{24}=-0{.}0093\in[-0{.}0347,0{.}0373]$ is close to zero and with probability 95\% belongs to an interval containing zero and approximately centered around zero. We may also notice that the three-wise dependence $\mu_{134}=-0{.}1252$ is quite similar to the four-wise dependence $\mu_{1234}=-0{.1254},$ so we may discard \texttt{crcard} as a relevant variable to predict \texttt{exited}, and therefore fit a reduced trivariate MBD with the remaining variables, renumbering them as follows:
\begin{enumerate}
    \item \texttt{gender}: 1=male, 0=female.
    \item \texttt{active}: customer is an active member (1=yes, 0=no).
    \item \texttt{exited}: customer has churned (1=yes, 0=no).
\end{enumerate}

In this case $n=3$ and therefore $2^{n}=8$ probabilities to estimate, $2^{n}-1=7$ parameters $\theta_{\!A}$ to estimate, and $2^{n}-(n+1)=4$ multivariate dependencies $\mu_A$ to estimate (pairwise and three-wise). Table \ref{tab:ex5B} summarizes point and 95\% probability interval estimations for the probabilities $p_{ijk},$ and tables \ref{tab:ex5C} and \ref{tab:ex5D} similar estimations for the $\theta_{\!A}$ parameters and $\mu_A$ dependencies, respectively.

\begin{table}[h]
    \begin{tabular}{|c|c|c|c|} \hline
    {       } & $p_{ijk}$ lower & $p_{ijk}$ median & $p_{ijk}$ upper  \\
    $(i,j,k)$ & quantile $0{.}025$ & quantile $0{.}500$ & quantile $0{.}975$ \\ \hline\hline  
    $(0,0,0)$ & 0.1465 & 0.1535 & 0.1606 \\
    $(0,0,1)$ & 0.0675 & 0.0725 & 0.0777 \\
    $(0,1,0)$ & 0.1793 & 0.1869 & 0.1946 \\
    $(0,1,1)$ & 0.0376 & 0.0414 & 0.0454 \\
    $(1,0,0)$ & 0.1935 & 0.2013 & 0.2092 \\
    $(1,0,1)$ & 0.0533 & 0.0578 & 0.0625 \\
    $(1,1,0)$ & 0.2459 & 0.2544 & 0.2630 \\
    $(1,1,1)$ & 0.0288 & 0.0321 & 0.0357 \\ \hline 
    \end{tabular}
    \caption{Point and 95\% probability interval estimations for the probabilities $p_{ijk}$ in the trivariate MBD model for the binary variables (\texttt{gender,active,exited}) from  bank churn data in \cite{Meshram}. The Julia code to generate the results is in Appendix \ref{secA3}.} 
    \label{tab:ex5B}
\end{table}

\begin{table}[h]
	\begin{tabular}{|c|c|c|c|} \hline
		{        }             & $\theta_{\!A}$ lower & $\theta_{\!A}$ median & $\theta_{\!A}$ upper  \\
		$A\subseteq\{1,2,3\}$  & quantile $0{.}025$ & quantile $0{.}500$ & quantile $0{.}975$ \\ \hline\hline  
		$\{1,2,3\}$            & 0.1465 & 0.1535 & 0.1606 \\
		$\{1,2\}$              & 0.2179 & 0.2260 & 0.2343 \\
		$\{1,3\}$              & 0.3311 & 0.3404 & 0.3497 \\
		$\{2,3\}$              & 0.3454 & 0.3548 & 0.3642 \\ 
		$\{1\}$                & 0.4445 & 0.4543 & 0.4641 \\
		$\{2\}$                & 0.4753 & 0.4851 & 0.4949 \\ 
		$\{3\}$                & 0.7881 & 0.7961 & 0.8039 \\ \hline 
	\end{tabular}
	\caption{Point and 95\% probability interval estimations for the parameters $\theta_{\!A}$ in the trivariate MBD model for the binary variables (\texttt{gender,active,exited}) from  bank churn data in \cite{Meshram}. The Julia code to generate the results is in Appendix \ref{secA3}.} 
	\label{tab:ex5C}
\end{table}

\begin{table}[h]
	\begin{tabular}{|c|r|r|r|} \hline
		{       }              & $\mu_A$ lower & $\mu_A$ median & $\mu_A$ upper  \\
		$A\subseteq\{1,2,3\}$  & quantile $0{.}025$ & quantile $0{.}500$ & quantile $0{.}975$ \\ \hline\hline  
		$\{1,2,3\}$            & -0.1537 & -0.1252 & -0.0966 \\
		$\{1,2\}$              &  0.0032 &  0.0240 &  0.0449 \\ 
		$\{1,3\}$              & -0.2270 & -0.1916 & -0.1561 \\  
		$\{2,3\}$              & -0.3359 & -0.2994 & -0.2625 \\ \hline 
	\end{tabular}
	\caption{Point and 95\% probability interval estimations for the dependencies $\mu_{A}$ in the trivariate MBD model for the binary variables (\texttt{gender,active,exited}) from  bank churn data in \cite{Meshram}. The Julia code to generate the results is in Appendix \ref{secA3}.} 
	\label{tab:ex5D}
\end{table}

With the trivariate fitted MBD model we may proceed to calculate unconditional and conditional probabilities for the main variable of interest \texttt{exited}, as summarized in Table \ref{tab:ex5E}. With such probabilities we define 5 rules for prediction, and then we estimate their accuracy through simulations:
\begin{itemize}
    \item \textit{Rule 1:} Non conditional simulation of 10,000 observations from a Bernoulli random variable with parameter $\mathbb{P}(\text{\texttt{exited}}=1)=0{.}2039$ (case 1 in Table \ref{tab:ex5E}).
    \item \textit{Rule 2:} Conditional simulation of 10,000 observations, depending on the value of \texttt{gender} in each of the 10,000 registries of the churn data, from a Bernoulli with parameter $\mathbb{P}(\text{\texttt{exited}}=1\,|\,\text{\texttt{gender}}=1)=0{.}1648$ or $\mathbb{P}(\text{\texttt{exited}}=1\,|\,\text{\texttt{gender}}=0)=0{.}2508,$ see cases 2 and 3 in Table \ref{tab:ex5E}.
    \item \textit{Rule 3:} Conditional simulation of 10,000 observations, depending on the value of \texttt{active} in each of the 10,000 registries of the churn data, from a Bernoulli with parameter according with cases 4 and 5 in Table \ref{tab:ex5E}.
    \item \textit{Rule 4:} Conditional simulation of 10,000 observations, depending on the value of \texttt{gender} and \texttt{active} in each of the 10,000 registries of the churn data, from a Bernoulli with parameter accordingly with cases 6, 7, 8, and 9, in Table \ref{tab:ex5E}.
    \item \textit{Rule 5:} Similar to \textit{Rule 4} but fitting a trivariate MBD such that the bivariate dependencies are the same as in Table \ref{tab:ex5D}, but setting the trivariate dependence parameter $\mu_{123}=0$ instead of $-0{.}1252.$
\end{itemize}

The purpose of Rule 5 is to compare its accuracy versus Rule 4 and notice the effect of ignoring dependencies beyond pairwise ones, in this case $\mu_{123}.$ For testing Rule 5 we need a vector of 8 trivariate probabilities $\{p_{ijk}:i,j,k\in{0,1}\}$ such that the marginals are the same (that is, same $\theta_1=0{.}4543,$ $\theta_2=0{.}4851,$ and $\theta_3=0{.}7961$ values as in Table \ref{tab:ex5C}), the same bivariate dependencies (that is, same $\mu_{12}=0{.}0240,$ $\mu_{13}=-0{.}1916,$ and $\mu_{23}=-0{.}2994$ values as in Table \ref{tab:ex5D}), but such that there is no trivariate dependence, that is $\mu_{123}=0.$ 

From (\ref{depmeas3d}) we notice that $\mu_{123}=0$ if and only if $\theta_{123}=\theta_1\theta_2\theta_3.$ Since $\mu_{12}=0.0240>0$ but $\mu_{13}=-0{.}1916<0$ and $\mu_{23}=-0{.}2994$ then from (\ref{depmeas2d}) we may obtain:
\begin{eqnarray*}
    \theta_{12} &=& (\min\{\theta_1,\theta_2\}-\theta_1\theta_2)\mu_{12} \,+\, \theta_1\theta_2\,, \\
    \theta_{13} &=& (\theta_1\theta_3-\max\{\theta_1+\theta_3+1,0\})\mu_{13} \,+\, \theta_1\theta_3\,, \\ 
    \theta_{23} &=& (\theta_2\theta_3-\max\{\theta_2+\theta_3+1,0\})\mu_{23} \,+\, \theta_2\theta_3\,.
\end{eqnarray*}

With the above new values $\theta_{\!A}$ we may apply formulas (\ref{p000})-(\ref{p111}) to obtain the required probabilities for Rule 5. Each of the 5 rules was applied 1 million times to each of the 10,000 registries of the churn data, and for each repetition the percentage of correct predictions compared to the real observed value of \texttt{exited} was calculated, and finally then averaged. The results are summarized in Table \ref{tab:ex5F}, from which the following remarks are derived:
\begin{itemize}
    \item Since there are significant bivariate dependencies $\mu_{13}\equiv\mu(\text{\texttt{gender}},\text{\texttt{exited}})=-0{.}1916$ and $\mu_{23}\equiv\mu(\text{\texttt{active}},\text{\texttt{exited}})=-0{.}2994,$ see Table \ref{tab:ex5D}, it was expected an increase of accuracy in predicting \texttt{exited} by just conditioning in one of the variables \texttt{gender} (Rule 2) or \texttt{active} (Rule 3) with respect to not conditioning at all (Rule 1), and the accuracy of Rule 3 is larger than Rule 2 since $|\mu_{23}|>|\mu_{13}|.$
    \item Since both variables \texttt{gender} and \texttt{active} have some predictive power for \texttt{exited}, according to the associated bivariate dependencies $\mu_{13}$ and $\mu_{23},$ it was expected to have even more accuracy by conditioning in both variables simultaneously, so there is no surprise that Rules 4 and 5 have better accuracies than Rules 2 and 3. 
    \item Rule 4 has better accuracy than Rule 5, as expected, because the latter ignores the trivariate dependence $\mu_{123}=-0{.}1252.$
\end{itemize}

\begin{table}[h]
	\begin{tabular}{|c|c|c|} \hline
		case & condition & $\mathbb{P}(\text{\texttt{exited}}=1\,|\,\text{condition})$ \\ \hline\hline  
		1 & none & 0.2039 \\
		2 & \texttt{gender} = 1 & 0.1648 \\  
        3 & \texttt{gender} = 0 & 0.2508 \\ 
        4 & \texttt{active} = 1 & 0.1428 \\ 
        5 & \texttt{active} = 0 & 0.2686 \\ 
        6 & \texttt{gender} = 1 and \texttt{active} = 1 & 0.1120 \\ 
        7 & \texttt{gender} = 1 and \texttt{active} = 0 & 0.2231 \\ 
        8 & \texttt{gender} = 0 and \texttt{active} = 1 & 0.1814 \\ 
        9 & \texttt{gender} = 0 and \texttt{active} = 0 & 0.3208 \\ \hline
	\end{tabular}
	\caption{Probabilities for the main variable of interest \texttt{exited} under different conditions, in the trivariate MBD model for the binary variables (\texttt{gender,active,exited}) from  bank churn data in \cite{Meshram}. The Julia code to generate the results is in Appendix \ref{secA3}.} 
	\label{tab:ex5E}
\end{table}

\begin{table}[h]
	\begin{tabular}{|c|c|} \hline
		Rule \# & Accuracy \\ \hline\hline  
		1 & 67.54\% \\
		2 & 67.91\% \\  
        3 & 68.33\% \\ 
        \textbf{4} & \textbf{68.69}\% \\ 
        5 & 68.56\% \\ \hline
	\end{tabular}
	\caption{Accuracy of predictions for \texttt{exited} according to the 5 defined rules in the trivariate MBD model for the binary variables (\texttt{gender,active,exited}) from  bank churn data in \cite{Meshram}. The Julia code to generate the results is in Appendix \ref{secA3}.} 
	\label{tab:ex5F}
\end{table}

\subsection{A real data example: COVID-19 in Mexico}

Data from official statistics about COVID-19 in Mexico during the first year of the pandemic, available for download from \cite{bib17} as a \texttt{covid2020.csv} file, with 2.15 million patient registries and 15 binary variables measured (among many others available), with the following description of the variables:

\begin{enumerate}
	\item \texttt{SEXO} = 1 for male, 0 for female.
	\item \texttt{TIPO\_PACIENTE} = 1 for hospitalized, 0 non-hospitalized.
	\item \texttt{DIABETES} = 1 for diabetic, 0 if not.
	\item \texttt{EPOC} = 1 for chronic obstructive pulmonary disease (COPD), 0 if not.
	\item \texttt{INMUSUPR} = 1 for immunosuppression, 0 if not.
	\item \texttt{HIPERTENSION} = 1 for hypertension, 0 if not.
	\item \texttt{CARDIOVASCULAR} = 1 for cardiovascular disease, 0 if not.
	\item \texttt{OBESIDAD} = 1 for obesity, 0 if not.
	\item \texttt{RENAL\_CRONICA} = 1 for chronic kidney disease (CKD), 0 if not.
	\item \texttt{TABAQUISMO} = 1 for smoker, 0 for non-smoker.
	\item \texttt{e00} = 1 for ages between 0 and 19, 0 if not.
	\item \texttt{e20} = 1 for ages between 20 and 39, 0 if not.
	\item \texttt{e40} = 1 for ages between 40 and 64, 0 if not.
	\item \texttt{e65} = 1 for ages between 65 and older, 0 if not.
	\item \texttt{MUERTE} = 1 for death, 0 if survived.
\end{enumerate}

In this case $n=15$ and therefore there are $2^{n}-1=32,768$ probabilities to estimate, $2^{n}-1$ parameters $\theta_A$ to estimate, and $2^{n}-(n+1)$ multivariate dependencies $\mu_A$ to estimate (pairwise, three-wise, up to 15-wise). The Julia code to fit a MBD is in Appendix \ref{secA4}. 

In Table \ref{tab:ex4} it is calculated the 20 largest values for dependence parameters that include variable 15 (death), from which it is possible to identify in all cases the presence of variable 2 (hospitalization), and in several cases variables 4, 9, 5, 7 and 14 (COPD, CKD, immunosuppression, cardiovascular disease, and age 65 or older). 

\begin{table}[h]
	\begin{tabular}{|c|c|} \hline
		{       }              & {point estimate}  \\
		$A\subseteq\{1,\ldots,15\}$  & $\mu_A$ \\ \hline\hline  
		$\{2,15\}$                   & 0.9021 \\
		$\{2,4,15\}$                 & 0.8590 \\ 
		$\{2,9,15\}$                 & 0.8581 \\ 
		$\{2,5,15\}$                 & 0.8550 \\ 
		$\{2,4,9,15\}$				 & 0.8311 \\
		$\{2,7,15\}$				 & 0.8310 \\
		$\{2,5,9,15\}$				 & 0.8283 \\
		$\{2,4,5,15\}$				 & 0.8272 \\
		$\{2,7,9,15\}$				 & 0.8094 \\
		$\{2,4,5,9,15\}$			 & 0.8090 \\
		$\{2,4,7,15\}$				 & 0.8088 \\
		$\{2,5,7,15\}$				 & 0.8032 \\
		$\{2,4,7,9,15\}$			 & 0.7940 \\
		$\{2,5,7,9,15\}$			 & 0.7897 \\
		$\{2,4,5,7,15\}$			 & 0.7879 \\ 
		$\{2,4,5,7,9,15\}$			 & 0.7784 \\ 
		$\{2,4,9,14,15\}$			 & 0.6594 \\
		$\{2,4,14,15\}$				 & 0.6583 \\
		$\{2,14,15\}$				 & 0.6565 \\ 
		$\{2,9,14,15\}$				 & 0.6560 \\ \hline 
	\end{tabular}
	\caption{The 20 largest values for dependence parameters that include variable 15 (death). For details of the calculations see Appendix \ref{secA4}.} 
	\label{tab:ex4}
\end{table}

It was of main interest to identify the variables more closely related to the fatal outcome of death (variable 15). This is especially relevant because conditioning on such variables a better assessment of mortality risk could be obtained. In this data the (non-conditional) mortality rate was $10.1\%$ but conditioning on variables 2 (hospitalization), 4 (COPD), 9 (CKD), 5 (immunosuppression), 7 (cardiovascular) and 14 (age 65 and older), it increases up to $66.3\%$ (for details of the calculations see Appendix \ref{secA4}).

\section{Conclusions}

In this paper, a novel subcopula-based characterization of dependence for the MBD has been introduced. By leveraging Sklar's theorem, it has been shown that subcopulas can effectively capture the dependence structure between Bernoulli random variables, extending the utility of subcopula theory to binary data. The proposed approach provides a flexible framework for studying multivariate binary data by separating the marginal distributions from the joint dependence structure.

Explicit formulas and bounds for both bivariate and multivariate Bernoulli distributions in terms of subcopulas were derived, allowing for a more granular understanding of the dependence parameters. In particular, dependence measures were obtained for bivariate and trivariate Bernoulli distributions, and a general method is proposed to extend these results to higher dimensions. This work's findings confirm that subcopulas are a powerful tool for modeling and quantifying dependence of all orders in the MBD, not just pairwise.

%The proposed methodology not only advances the theoretical understanding of subcopulas but also opens up new possibilities for practical applications in fields dealing with multivariate binary data. Potential applications include, for example:
%\begin{itemize}
%	\item Epidemiology: Modeling the joint occurrence of comorbidities in disease outbreaks;
%	\item Genetics: Analyzing dependencies in gene expression data where outcomes are binary;
%	\item Finance: Assessing the joint probability of binary financial events, such as default occurrences;
%\end{itemize}
%and any other disciplines where understanding the dependence structure between binary variables is essential. Moreover, the bayesian inference framework presented offers a robust tool for parameter estimation and uncertainty quantification, enhancing the practical implementation of the proposed approach.

%Computational implementations of the MBD are quite scarce, so the set of functions provided in section \ref{JuliaMBD} and downloadable from \cite{bib17} is a contribution that may help in analyzing and simulating multivariate binary data. Future research could explore further generalizations of the subcopula approach to other types of discrete data and investigate its performance in applied contexts. Additionally, incorporating covariates or other forms of heterogeneity into the subcopula framework may provide more nuanced models, particularly for data sets with more complex dependence structures.

The proposed methodology not only advances the theoretical understanding of subcopulas but also opens up new possibilities for practical applications in fields dealing with multivariate binary data. For example, in epidemiology, the framework can model the joint occurrence of comorbidities in infectious disease outbreaks, such as diabetes, hypertension, COPD, and mortality in COVID-19 patients, or assess the co-occurrence of vaccine side effects. In genetics, it can be applied to analyze dependencies among binary outcomes such as gene activation/inactivation, the presence or absence of mutations, or epistatic interactions between SNPs (Single Nucleotide Polymorphisms). In finance, the approach is directly relevant for credit risk modeling, where default/no-default indicators across obligors must be analyzed jointly, as well as for systemic risk assessment in portfolios, where simultaneous threshold-crossing events in asset returns are of interest.

Moreover, the Bayesian inference framework presented here offers robust tools for parameter estimation and uncertainty quantification, enhancing the practical implementation of the proposed approach. Computational implementations of the MBD are quite scarce, so the set of functions provided in Section 6 and downloadable from \cite{bib17} represents a contribution that facilitates the analysis and simulation of multivariate binary data. Future research could explore further generalizations of the subcopula approach to other types of discrete data and investigate its performance in applied contexts. Additionally, incorporating covariates or other forms of heterogeneity into the subcopula framework may provide more nuanced models, particularly for data sets with more complex dependence structures.

%%%%%%%%% Response to Editor request for limitations
Although the proposed characterization is explicit and fully compatible, the parameter dimension grows as $2^n-1,$ so estimation can become data-hungry and computationally heavy for large values of $n.$ In such settings, practical use may require sparsity/regularization, structured restrictions, or focusing on selected subsets $A$ of interest.

\backmatter

\begin{appendices}

\section{Julia code for the examples}

All the source code is available in a public repository at \cite{bib17} for full reproducibility, and must be loaded and executed following the provided indications.

\subsection{A theoretical example (6.1)}\label{secA1}

\begin{verbatim}
	
	# setup parameters and MBD model
	pp = [0.15,0.21,0.21,0.03,0.21,0.03,0.03,0.13];
	X = MBerDep(pp);
	
	# parameters
	[X.dparam.idx X.dparam.value]
	
	# dependence measures
	[X.dmeas.idx X.dmeas.value]
	
\end{verbatim}

\subsection{A simulation example (6.2)}\label{secA2}

Define a trivariate Bernoulli distribution applying Algorithm 1 from Section \ref{sec:trivariate}:
\begin{verbatim}
    begin
    p000 = 0.1                                            
    p001, p010, p100 = 0.2, 0.1, 0.05                     
    p011, p101, p110 = 0.2, 0.15, 0.1                     
    p111 = 1 - sum([p000, p001, p010, p011, p100, p101, p110])  
    pp = [p000, p001, p010, p011, p100, p101, p110, p111] 
    X = MBerDep(pp)
    end    
\end{verbatim}

Simulate a random sample of size $3,000$:
\begin{verbatim}
    begin
    Random.seed!(1234) # for reproducibility
    simX = MBerSim(pp, 3_000);
    end
\end{verbatim}

Interval and point estimations for the probability parameters:
\begin{verbatim}
    infer = MBerInf(simX, prior = 1/2, nsim = 10_000, probint = 0.99);
    infer.probs # estimations
    [X.binprob.idx X.binprob.value] # theoretical probabilities
\end{verbatim}

Interval and point estimations for the dependence parameters:
\begin{verbatim}
    begin
    infer.dparam # estimations
    [X.dparam.idx X.dparam.value] # theoretical dependence parameters
    end
\end{verbatim}

Coverage rates for all parameters:
\begin{verbatim}
    @time begin # WARNING: 10_000 simulations take around 3.5 hours
      nsim = 10_000
      nobs = 3_000
      cover_probs = zeros(Bool, nsim, 8)
      cover_param = zeros(Bool, nsim, 7)
      cover_meas = zeros(Bool, nsim, 4)
      for i in 1:nsim
         simX = MBerSim(pp, nobs)
         infer = MBerInf(simX,prior=1/2,nsim=100000,probint=0.99)
         for j in 1:8
           cover_probs[i,j]=(infer.probs[j,2]<=pp[j]<=infer.probs[j,4])
         end
         for j in 1:7
           cover_param[i,j]=(infer.dparam[j,2]<=X.dparam.value[j]
           <=infer.dparam[j,4])
         end
         for j in 1:4
           cover_meas[i,j]=(infer.dmeas[j,2]<=X.dmeas.value[j]
           <=infer.dmeas[j,4])
         end
      end
      println("Coverage rates (99% intervals):")
      println("Probabilities: ", mean(cover_probs, dims = 1))
      println("Parameters:    ", mean(cover_param, dims = 1))
      println("Dependencies:  ", mean(cover_meas, dims = 1))
      println()
    end
\end{verbatim}

\subsection{A real data example: Bank churn data (6.3)}\label{secA3}

\begin{verbatim}
    begin # Read data as a dataframe
        println("Example Section 6.4: Bank churn data")
        df = CSV.read("churnbindata.csv", DataFrame)
        show(describe(df), allrows = true)
    end
\end{verbatim}

\begin{verbatim}
    begin # Convert dataframe to a matrix
        binmat = zeros(Int, size(df))
        for c in 1:ncol(df)
            binmat[:, c] = df[:, c]
        end
        binmat
    end
\end{verbatim}

\begin{verbatim}
    # Point and 95% probability interval estimations 
    @time inference4v = MBerInf(binmat, nsim = 1_000_000)
    inference4v.dmeas

    # Discard variable 'crcard' (column 2)
    binmat = binmat[:, [1,3,4]]
    @time inference3v = MBerInf(binmat, nsim = 1_000_000)

    # Point and 95% probability interval estimations for...
    inference3v.probs # ... probabilities
    inference3v.dparam # ... parameters
    inference3v.dmeas # ... dependencies
\end{verbatim}

\begin{verbatim}
    # Probabilities for exiting the bank
    begin
    condProb = DataFrame(gender = [missing,1,0,missing,missing,1,1,0,0],
                         active = [missing,missing,missing,1,0,1,0,1,0],
                         Pexit = zeros(9))
    condProb.Pexit[1] = 1 - inference3v.dparam[7,3]
    # warning due to rounding
    X123 = MBerDep(Float64.(inference3v.probs[:, 3]))
    condProb.Pexit[2] = MBerCond(X123.binprob.value, [3], [1],
    [1]).binprob.value[2]
    condProb.Pexit[3] = MBerCond(X123.binprob.value, [3], [1],
    [0]).binprob.value[2]
    condProb.Pexit[4] = MBerCond(X123.binprob.value, [3], [2],
    [1]).binprob.value[2]
    condProb.Pexit[5] = MBerCond(X123.binprob.value, [3], [2],
    [0]).binprob.value[2]  
    condProb.Pexit[6] = MBerCond(X123.binprob.value, [3], [1,2],
    [1,1]).binprob.value[2]
    condProb.Pexit[7] = MBerCond(X123.binprob.value, [3], [1,2],
    [1,0]).binprob.value[2]
    condProb.Pexit[8] = MBerCond(X123.binprob.value, [3], [1,2],
    [0,1]).binprob.value[2]
    condProb.Pexit[9] = MBerCond(X123.binprob.value, [3], [1,2],
    [0,0]).binprob.value[2]
    display(condProb)
    end
\end{verbatim}

\begin{verbatim}
# Prediction rules
function predNonCond(nsim, condProb)
    B = Bernoulli(condProb.Pexit[1])
    accuracy = 0.0
    for k in 1:nsim
        accuracy += mean(rand(B, 10_000) .== binmat[:, 3]) / nsim
    end
    return accuracy
end
function predGivenGender(nsim, condProb)
    ival1 = findall(binmat[:, 1] .== 1)
    ival0 = findall(binmat[:, 1] .== 0)
    n1 = length(ival1)
    n0 = length(ival0)
    vsim = fill(-99999999, 10_000)
    accuracy = 0.0
    for k in 1:nsim
        vsim[ival1] = rand(Bernoulli(condProb.Pexit[2]), n1)
        vsim[ival0] = rand(Bernoulli(condProb.Pexit[3]), n0)
        accuracy += mean(vsim .== binmat[:, 3]) / nsim
    end
    return accuracy
end
function predGivenActive(nsim, condProb)
    ival1 = findall(binmat[:, 2] .== 1)
    ival0 = findall(binmat[:, 2] .== 0)
    n1 = length(ival1)
    n0 = length(ival0)
    vsim = fill(-99999999, 10_000)
    accuracy = 0.0
    for k in 1:nsim
        vsim[ival1] = rand(Bernoulli(condProb.Pexit[4]), n1)
        vsim[ival0] = rand(Bernoulli(condProb.Pexit[5]), n0)
        accuracy += mean(vsim .== binmat[:, 3]) / nsim
    end
    return accuracy
end
function predGivenGenderActive(nsim, condProb)
    X1X2 = Vector{Int}[]
    for j in 1:10_000
        push!(X1X2, binmat[j, [1,2]])
    end
    ival11 = findall(X1X2 .== [[1,1]])
    ival10 = findall(X1X2 .== [[1,0]])
    ival01 = findall(X1X2 .== [[0,1]])
    ival00 = findall(X1X2 .== [[0,0]])
    n11 = length(ival11)
    n10 = length(ival10)
    n01 = length(ival01)
    n00 = length(ival00)
    vsim = fill(-99999999, 10_000)
    accuracy = 0.0
    for k in 1:nsim
        vsim[ival11] = rand(Bernoulli(condProb.Pexit[6]), n11)
        vsim[ival10] = rand(Bernoulli(condProb.Pexit[7]), n10)
        vsim[ival01] = rand(Bernoulli(condProb.Pexit[8]), n01)
        vsim[ival00] = rand(Bernoulli(condProb.Pexit[9]), n00)
        accuracy += mean(vsim .== binmat[:, 3]) / nsim
    end
    return accuracy
end
\end{verbatim}

\begin{verbatim}
# Accuracy for each rule 
nsim = 1_000_000 # 4 minutes approx 
@time predict=DataFrame(rule=["Non-conditional","Given gender",
                      "Given active", "Given gender & active"],
            accuracy = [predNonCond(nsim, condProb), 
                        predGivenGender(nsim, condProb),
                        predGivenActive(nsim, condProb), 
                        predGivenGenderActive(nsim, condProb)])
\end{verbatim}

\begin{verbatim}
# Accuracy given gender and active status
# but setting trivariate dependence to zero 
begin
    th1 = inference3v.dparam[5,3]
    th2 = inference3v.dparam[6,3]
    th3 = inference3v.dparam[7,3]
    th123 = th1 * th2 * th3 
    mu12 = inference3v.dmeas[2,3]
    mu13 = inference3v.dmeas[3,3]
    mu23 = inference3v.dmeas[4,3]
    th12 = (min(th1,th2) - th1*th2)*mu12 + th1*th2
    th13 = (th1*th3 - max(th1+th3-1,0))*mu13 + th1*th3
    th23 = (th2*th3 - max(th2+th3-1,0))*mu23 + th2*th3
    p000 = th123 
    p001 = th12 - th123
    p010 = th13 - th123
    p100 = th23 - th123
    p011 = th1 - th12 - th13 + th123
    p101 = th2 - th12 - th23 + th123
    p110 = th3 - th13 - th23 + th123
    p111 = 1 - (p000 + p001 + p010 + p011 + p100 + p101 + p110)
    pp = [p000, p001, p010, p011, p100, p101, p110, p111]
    Xzero3dep = MBerDep(pp)
end;
# dependence measures without and with zero trivariate dependence
[inference3v.dmeas[:, [1,3]] Xzero3dep.dmeas.value]
\end{verbatim}

\begin{verbatim}
# Probabilities for exiting the bank with zero trivariate dependence
begin
condProb3 = DataFrame(gender = [missing,1,0,missing,missing,1,1,0,0],
                      active = [missing,missing,missing,1,0,1,0,1,0],
                      Pexit = zeros(9))
condProb3.Pexit[1] = 1 - Xzero3dep.dparam.value[7]
condProb3.Pexit[2] = MBerCond(Xzero3dep.binprob.value, [3], [1],
[1]).binprob.value[2]
condProb3.Pexit[3] = MBerCond(Xzero3dep.binprob.value, [3], [1],
[0]).binprob.value[2]
condProb3.Pexit[4] = MBerCond(Xzero3dep.binprob.value, [3], [2],
[1]).binprob.value[2]
condProb3.Pexit[5] = MBerCond(Xzero3dep.binprob.value, [3], [2],
[0]).binprob.value[2]  
condProb3.Pexit[6] = MBerCond(Xzero3dep.binprob.value, [3], [1,2],
[1,1]).binprob.value[2]
condProb3.Pexit[7] = MBerCond(Xzero3dep.binprob.value, [3], [1,2],
[1,0]).binprob.value[2]
condProb3.Pexit[8] = MBerCond(Xzero3dep.binprob.value, [3], [1,2],
[0,1]).binprob.value[2]
condProb3.Pexit[9] = MBerCond(Xzero3dep.binprob.value, [3], [1,2],
[0,0]).binprob.value[2]
display(condProb3)
end
\end{verbatim}

\begin{verbatim}
# Accuracy for each rule setting trivariate dependence to zero
nsim = 1_000_000 # 4 minutes approx 
@time predict3=DataFrame(rule=["Non-conditional","Given gender",
                         "Given active","Given gender & active"],
                accuracy = [predNonCond(nsim, condProb3), 
                        predGivenGender(nsim, condProb3),
                        predGivenActive(nsim, condProb3), 
                        predGivenGenderActive(nsim, condProb3)]);
\end{verbatim}

\begin{verbatim}
# Comparison of accuracies with and without trivariate dependence
println("Accuracy with trivariate dependence:")
predict
println("Accuracy with zero trivariate dependence:")
predict3
\end{verbatim}

\subsection{A real data example: COVID-19 in Mexico (6.4)}\label{secA4}

\begin{verbatim}	
	begin # load data
		df = CSV.read("covid2020.csv", DataFrame)
		show(describe(df), allrows = true)
		data = zeros(Int, size(df))
		for c in 1:ncol(df)
			data[:, c] = df[:, c]
		end
		println()
		data
	end
\end{verbatim}

\begin{verbatim}	
@time estim = MBerEst(data); # Fit MBD model (5 minutes approx)	
\end{verbatim}

\begin{verbatim}
# 20 largest values for dependence parameters
# that include variable 15 (death)
begin 
iord = sortperm(estim.dmeas.value, rev = true)
midx = estim.dmeas.idx[iord]
m = estim.dmeas.value[iord]
iMue = findall(x -> 15 in x, midx)[midx[iMue] m[iMue]][1:20, :]
end
\end{verbatim}

\begin{verbatim}	
begin # conditional risk
    pd = mean(data[:, 15]) # general mortality rate
    estim2 = MBerEst(data[:, [2,4,5,7,9,14,15]])
    # conditional mortality rate
    pdcond = MBerCond(estim2.binprob.value, [7], [1,2,3,4,5,6],	
	       [1,1,1,1,1,1]).binprob.dic[[1]]
    println("P(death) = ", pd)
    println("P(death|hospital, COPD, immunosup, cardio,
	      CKD, age 65+) = ", pdcond)
end
\end{verbatim}

\end{appendices}

\section*{DECLARATIONS} 

\subsection*{Ethics approval and consent to participate}
Not applicable. This study did not involve human participants, animals, or sensitive personal data requiring ethical approval.

\subsection*{Consent for publication}
Not applicable. This manuscript does not contain data from any individual person.

\subsection*{Availability of data and materials}
All the data and programming code is openly available at the following Github repository where the author of this repository is the author of this manuscript: \url{https://github.com/aerdely/MultivariateBernoulli}

\subsection*{Competing interests}
The author declares no competing interests as defined by Springer, or other interests that might be perceived to influence the results and/or discussion reported in this paper.

\subsection*{Funding}
Direcci\'{o}n General de Asuntos del Personal Acad\'{e}mico, Universidad Nacional Aut\'{o}noma de M\'{e}xico, PAPIIT IN104425.

\subsection*{Authors' contributions}
As a single author A.E has worked all the details in this contribution.

\subsection*{Acknowledgements}
The author gratefully acknowledges the reviewers and the Editor whose comments helped to significantly improve this work.

\bibliography{sn-bibliography}% common bib file

\end{document}